\documentclass[screen=true,review=false,authorversion=true,sigplan]{acmart}
\renewcommand{\documentclass}[2][]{}
\newif\ifediting\editingfalse
\newif\iflong\longfalse
\documentclass[screen=true,review=true,sigplan]{acmart}

\makeatletter
\@ifundefined{ifediting}{
  \newif\ifediting%
  \editingtrue%
}{}
\makeatother

\makeatletter
\@ifundefined{iflong}{
  \newif\iflong%
  \longfalse%
}{}
\makeatother

\usepackage[utf8]{inputenc}
\usepackage{etoolbox}
\usepackage[capitalise,nameinlink,noabbrev]{cleveref}
\usepackage{paralist}
\usepackage[zi4]{inconsolata-slanted} 
\usepackage{needspace}
\usepackage{flushend}

\usepackage{xspace}
\newcommand*{\cn}[1][]{%
  \textcolor{red}{\textbf{%
    (?!%
    \ifx\relax#1\relax\else{} #1\fi
    )%
  }}\xspace
}

\usepackage{draft-todo}
\ifediting
  \drafttrue

  \paperwidth=\dimexpr\paperwidth+6cm\relax%
  \oddsidemargin=\dimexpr\oddsidemargin+3cm\relax%
  \evensidemargin=\dimexpr\evensidemargin+3cm\relax%
  \marginparsep=\dimexpr\marginparsep+0.8cm\relax%
  \marginparwidth=\dimexpr\marginparwidth+2.6cm\relax%
\else
  \draftfalse
  \renewcommand{\todo}[2][]{}
\fi

\newtodo{asz}{violet!25}        
\newtodo{jb}{green!40}          
\newtodo{criz}{blue!67!cyan!20} 
\newtodo{scw}{red!20}           
\newtodo{leo}{yellow!67}        
\newtodo{jp}{teal!50}           

\DeclareUnicodeCharacter{2218}{\circ}

\makeatletter
\@ifundefined{lhs2tex.lhs2tex.sty.read}%
  {\@namedef{lhs2tex.lhs2tex.sty.read}{}%
   \newcommand\SkipToFmtEnd{}%
   \newcommand\EndFmtInput{}%
   \long\def\SkipToFmtEnd#1\EndFmtInput{}%
  }\SkipToFmtEnd

\newcommand\ReadOnlyOnce[1]{\@ifundefined{#1}{\@namedef{#1}{}}\SkipToFmtEnd}
\usepackage{amstext}
\usepackage{amssymb}
\usepackage{stmaryrd}
\DeclareFontFamily{OT1}{cmtex}{}
\DeclareFontShape{OT1}{cmtex}{m}{n}
  {<5><6><7><8>cmtex8
   <9>cmtex9
   <10><10.95><12><14.4><17.28><20.74><24.88>cmtex10}{}
\DeclareFontShape{OT1}{cmtex}{m}{it}
  {<-> ssub * cmtt/m/it}{}

\DeclareFontShape{OT1}{cmtt}{bx}{n}
  {<5><6><7><8>cmtt8
   <9>cmbtt9
   <10><10.95><12><14.4><17.28><20.74><24.88>cmbtt10}{}
\DeclareFontShape{OT1}{cmtex}{bx}{n}
  {<-> ssub * cmtt/bx/n}{}

\newcommand{\Conid}[1]{\mathit{#1}}
\newcommand{\Varid}[1]{\mathit{#1}}
\newcommand{\anonymous}{\kern0.06em \vbox{\hrule\@width.5em}}

\usepackage{polytable}

\@ifundefined{mathindent}%
  {\newdimen\mathindent\mathindent\leftmargini}%
  {}%

\def\resethooks{%
  \global\let\SaveRestoreHook\empty
  \global\let\ColumnHook\empty}
\newcommand*{\savecolumns}[1][default]%
  {\g@addto@macro\SaveRestoreHook{\savecolumns[#1]}}
\newcommand*{\restorecolumns}[1][default]%
  {\g@addto@macro\SaveRestoreHook{\restorecolumns[#1]}}
\newcommand*{\aligncolumn}[2]%
  {\g@addto@macro\ColumnHook{\column{#1}{#2}}}

\resethooks

\newcommand{\onelinecommentchars}{\quad-{}- }
\newcommand{\commentbeginchars}{\enskip\{-}
\newcommand{\commentendchars}{-\}\enskip}

\newcommand{\visiblecomments}{%
  \let\onelinecomment=\onelinecommentchars
  \let\commentbegin=\commentbeginchars
  \let\commentend=\commentendchars}

\newcommand{\invisiblecomments}{%
  \let\onelinecomment=\empty
  \let\commentbegin=\empty
  \let\commentend=\empty}

\visiblecomments

\newlength{\blanklineskip}
\newcommand{\hsindent}[1]{\quad}
\let\hspre\empty
\let\hspost\empty

\EndFmtInput
\makeatother
\ReadOnlyOnce{polycode.fmt}%
\makeatletter

\newcommand{\hsnewpar}[1]%
  {{\parskip=0pt\parindent=0pt\par\vskip #1\noindent}}

\newcommand{\hscodestyle}{}

\newcommand{\sethscode}[1]%
  {\expandafter\let\expandafter\hscode\csname #1\endcsname
   \expandafter\let\expandafter\endhscode\csname end#1\endcsname}

  {\par\noindent
   \advance\leftskip\mathindent
   \hscodestyle
   \let\\=\@normalcr
   \let\hspre\(\let\hspost\)%
   \pboxed}%
  {\endpboxed\)%
   \par\noindent
   \ignorespacesafterend}

  {\hsnewpar\abovedisplayskip
   \advance\leftskip\mathindent
   \hscodestyle
   \let\hspre\(\let\hspost\)%
   \pboxed}%
  {\endpboxed%
   \hsnewpar\belowdisplayskip
   \ignorespacesafterend}

  {\hsnewpar\abovedisplayskip
   \advance\leftskip\mathindent
   \hscodestyle
   \let\\=\@normalcr
   \(\pboxed}%
  {\endpboxed\)%
   \hsnewpar\belowdisplayskip
   \ignorespacesafterend}

\newcommand{\plainhs}{\sethscode{plainhscode}}

\plainhs

  {\hsnewpar\abovedisplayskip
   \advance\leftskip\mathindent
   \hscodestyle
   \let\\=\@normalcr
   \(\parray}%
  {\endparray\)%
   \hsnewpar\belowdisplayskip
   \ignorespacesafterend}

  {\parray}{\endparray}

  {\(\parray}{\endparray\)}

\def\codeframewidth{\arrayrulewidth}
\RequirePackage{calc}

  {\parskip=\abovedisplayskip\par\noindent
   \hscodestyle
   \arrayrulewidth=\codeframewidth
   \tabular{@{}|p{\linewidth-2\arraycolsep-2\arrayrulewidth-2pt}|@{}}%
   \hline\framedhslinecorrect\\{-1.5ex}%
   \let\endoflinesave=\\
   \let\\=\@normalcr
   \(\pboxed}%
  {\endpboxed\)%
   \framedhslinecorrect\endoflinesave{.5ex}\hline
   \endtabular
   \parskip=\belowdisplayskip\par\noindent
   \ignorespacesafterend}

\newcommand{\framedhslinecorrect}[2]%
  {#1[#2]}

  {\(\def\column##1##2{}%
   \let\>\undefined\let\<\undefined\let\\\undefined
   \newcommand\>[1][]{}\newcommand\<[1][]{}\newcommand\\[1][]{}%
   \def\fromto##1##2##3{##3}%
   }{\) }%

  {\let\orighscode=\hscode
   \let\origendhscode=\endhscode
   \def\endhscode{\def\hscode{\endgroup\def\@currenvir{hscode}\\}\begingroup}
   \orighscode\def\hscode{\endgroup\def\@currenvir{hscode}}}%
  {\origendhscode
   \global\let\hscode=\orighscode
   \global\let\endhscode=\origendhscode}%

\makeatother
\EndFmtInput
\newenvironment{myhscode}%
  {\hsnewpar\abovedisplayskip
   \advance\leftskip\mathindent
   \hscodestyle
   \let\hspre\(\let\hspost\)%
   \pboxed}%
  {\endpboxed%
   \hsnewpar\belowdisplayskip
   \ignorespacesafterend}

\sethscode{myhscode}

\newcommand{\codefont}{\ttfamily}

\newcommand*{\EditName}[1]{\texttt{\textbf{#1}}}

\let\Varid\GenericID
\let\Conid\GenericID

\renewcommand{\onelinecomment}{\,--\hspace{.5pt}--\itshape\codefont{} }
\setcopyright{acmlicensed}
\acmPrice{15.00}
\acmDOI{10.1145/3167092}
\acmYear{2018}
\copyrightyear{2018}
\acmISBN{978-1-4503-5586-5/18/01}
\acmConference[CPP'18]{7th ACM SIGPLAN International Conference on Certified Programs and Proofs}{January 8--9, 2018}{Los Angeles, CA, USA}



\begin{document}

\title{Total Haskell is Reasonable Coq}

\author{Antal Spector-Zabusky \quad Joachim Breitner \quad Christine Rizkallah \quad Stephanie Weirich}
\email{{antals,joachim,criz,sweirich}@cis.upenn.edu}
\affiliation{%
  \institution{University of Pennsylvania}
  \streetaddress{3330 Walnut St}
  \city{Philadelphia}
  \state{PA}
  \postcode{19104}
  \country{USA}
}

\begin{abstract}
  We would like to use the Coq proof assistant to mechanically verify properties
  of Haskell programs.
  To that end, we present a tool, \jp{Either remove comma before `named', or
    remove `named' itself}named \texttt{hs-to-coq}, that translates
  total Haskell programs into Coq programs via a shallow embedding.
  We apply our tool in three case studies -- a lawful \ensuremath{\Conid{Monad}} instance,
  ``Hutton's razor'', and an existing data structure library -- and prove their
  correctness.
  These examples show that this approach is viable: both that \texttt{hs-to-coq}
  applies to existing Haskell code, and that the output it produces is amenable
  to verification.
\end{abstract}

\renewcommand{\authors}{Antal Spector-Zabusky, Joachim Breitner, Christine Rizkallah, and Stephanie Weirich}
\begingroup\sloppy
\maketitle
\fussy\endgroup
\renewcommand{\shortauthors}{Spector-Zabusky, Breitner, Rizkallah, and Weirich}

\section{Introduction}
\label{sec:introduction}

The Haskell programming language is a great tool for producing pure,
functional programs. Its type system tracks the use of impure features, such
as mutation and IO, and its standard library promotes the use of
mathematically-inspired structures that have strong algebraic properties. At
the same time, Haskell development is backed by an industrial-strength
compiler (the Glasgow Haskell Compiler, GHC)~\cite{GHC}, and supported by mature software development tools,
such as IDEs and testing environments.

\vfill\eject

However, Haskell programmers typically \emph{reason} about their code only
informally. Most proofs are done on paper, by hand, which is tedious,
error-prone, and does not scale.

On the other hand, the Coq proof assistant~\cite{Coq:manual} is a great tool for
writing proofs. It allows programmers to reason about total functional programs
conveniently, efficiently, and with high confidence. However, Coq lacks GHC's
extensive ecosystem for program development.

Therefore, we propose a multimodal approach to the verification of total
functional programs: write code in Haskell and prove it correct in Coq. To
support this plan, we have developed an automatic translator, called
\texttt{hs-to-coq}, that allows this approach to scale.

For example, consider the standard \ensuremath{\Varid{map}} function on lists (from the Haskell
Prelude), and the list \ensuremath{\Conid{Functor}} instance.
\begin{hscode}\SaveRestoreHook
\column{B}{@{}>{\hspre}l<{\hspost}@{}}%
\column{15}{@{}>{\hspre}l<{\hspost}@{}}%
\column{E}{@{}>{\hspre}l<{\hspost}@{}}%
\>[B]{}\Varid{map}\mathbin{::}(\Varid{a}\mathbin{\text{\codefont ->}}\Varid{b})\mathbin{\text{\codefont ->}}[\mskip1.5mu \Varid{a}\mskip1.5mu]\mathbin{\text{\codefont ->}}[\mskip1.5mu \Varid{b}\mskip1.5mu]{}\<[E]%
\\[-0.3ex]%
\>[B]{}\Varid{map}\;\Varid{f}\;[\mskip1.5mu \mskip1.5mu]{}\<[15]%
\>[15]{}\mathrel{=}[\mskip1.5mu \mskip1.5mu]{}\<[E]%
\\[-0.3ex]%
\>[B]{}\Varid{map}\;\Varid{f}\;(\Varid{x}\mathbin{:}\Varid{xs}){}\<[15]%
\>[15]{}\mathrel{=}\Varid{f}\;\Varid{x}\mathbin{:}\Varid{map}\;\Varid{f}\;\Varid{xs}{}\<[E]%
\\[\blanklineskip]%
\>[B]{}\textbf{\codefont instance}\;\Conid{Functor}\;[\mskip1.5mu \mskip1.5mu]\;\textbf{\codefont where}\;\Varid{fmap}\mathrel{=}\Varid{map}{}\<[E]%
\ColumnHook
\end{hscode}\resethooks
Our tool translates this Haskell program automatically to the analogous Coq
definitions.  The \ensuremath{\Varid{map}} function becomes the expected fixpoint.
\begin{hscode}\SaveRestoreHook
\column{B}{@{}>{\hspre}l<{\hspost}@{}}%
\column{3}{@{}>{\hspre}l<{\hspost}@{}}%
\column{9}{@{}>{\hspre}c<{\hspost}@{}}%
\column{9E}{@{}l@{}}%
\column{13}{@{}>{\hspre}l<{\hspost}@{}}%
\column{14}{@{}>{\hspre}l<{\hspost}@{}}%
\column{21}{@{}>{\hspre}l<{\hspost}@{}}%
\column{E}{@{}>{\hspre}l<{\hspost}@{}}%
\>[B]{}{\textbf{\codefont Definition}}\;\Varid{map}\;\mathord{\text{\codefont \{}}\Varid{a}\mathord{\text{\codefont \}}}\;\mathord{\text{\codefont \{}}\Varid{b}\mathord{\text{\codefont \}}}\mathbin{:}(\Varid{a}\mathbin{\text{\codefont ->}}\Varid{b})\mathbin{\text{\codefont ->}}\Varid{list}\;\Varid{a}\mathbin{\text{\codefont ->}}\Varid{list}\;\Varid{b}\mathbin{:=}{}\<[E]%
\\[-0.3ex]%
\>[B]{}\hsindent{3}{}\<[3]%
\>[3]{}{\textbf{\codefont fix}}\;\Varid{map}\;\Varid{arg\char95 62\char95 \char95 }\;\Varid{arg\char95 63\char95 \char95 }{}\<[E]%
\\[-0.3ex]%
\>[3]{}\hsindent{6}{}\<[9]%
\>[9]{}\mathbin{:=}{}\<[9E]%
\>[13]{}{\textbf{\codefont match}}\;\Varid{arg\char95 62\char95 \char95 },\Varid{arg\char95 63\char95 \char95 }\;{\textbf{\codefont with}}{}\<[E]%
\\[-0.3ex]%
\>[13]{}\hsindent{1}{}\<[14]%
\>[14]{}\mid \anonymous ,{}\<[21]%
\>[21]{}\Varid{nil}\mathbin{\text{\codefont =>}}\Varid{nil}{}\<[E]%
\\[-0.3ex]%
\>[13]{}\hsindent{1}{}\<[14]%
\>[14]{}\mid \Varid{f},{}\<[21]%
\>[21]{}\Varid{cons}\;\Varid{x}\;\Varid{xs}\mathbin{\text{\codefont =>}}\Varid{cons}\;(\Varid{f}\;\Varid{x})\;(\Varid{map}\;\Varid{f}\;\Varid{xs}){}\<[E]%
\\[-0.3ex]%
\>[13]{}{\textbf{\codefont end}}\mathop{\text{.}}{}\<[E]%
\ColumnHook
\end{hscode}\resethooks
Similarly, the \ensuremath{\Conid{Functor}} type class in Haskell turns into a Coq type class of
the same name, and Haskell's \ensuremath{\Conid{Functor}} instance for lists becomes a type class
instance on the Coq side.\jp{can you give this code?}

Once the Haskell definitions have been translated to Coq, users can prove
theorems about them.  For example, we provide a type class for \emph{lawful}
functors:
\begin{hscode}\SaveRestoreHook
\column{B}{@{}>{\hspre}l<{\hspost}@{}}%
\column{3}{@{}>{\hspre}c<{\hspost}@{}}%
\column{3E}{@{}l@{}}%
\column{6}{@{}>{\hspre}l<{\hspost}@{}}%
\column{9}{@{}>{\hspre}l<{\hspost}@{}}%
\column{12}{@{}>{\hspre}l<{\hspost}@{}}%
\column{E}{@{}>{\hspre}l<{\hspost}@{}}%
\>[B]{}{\textbf{\codefont Class}}\;\Conid{FunctorLaws}\;(\Varid{t}\mathbin{:}\Conid{Type}\mathbin{\text{\codefont ->}}\Conid{Type})\text{\phantom{x}}\text{\codefont\textasciigrave}\mathord{\text{\codefont \{}}\Conid{Functor}\;\Varid{t}\mathord{\text{\codefont \}}}\mathbin{:=}{}\<[E]%
\\[-0.3ex]%
\>[B]{}\hsindent{3}{}\<[3]%
\>[3]{}\mathord{\text{\codefont \{}}{}\<[3E]%
\>[6]{}\Varid{functor\char95 identity}\mathbin{:}{}\<[E]%
\\[-0.3ex]%
\>[6]{}\hsindent{3}{}\<[9]%
\>[9]{}{\textbf{\codefont forall}}\;\Varid{a}\;(\Varid{x}\mathbin{:}\Varid{t}\;\Varid{a}),\Varid{fmap}\;\Varid{id}\;\Varid{x}\mathrel{=}\Varid{x};{}\<[E]%
\\[-0.3ex]%
\>[6]{}\Varid{functor\char95 composition}\mathbin{:}{}\<[E]%
\\[-0.3ex]%
\>[6]{}\hsindent{3}{}\<[9]%
\>[9]{}{\textbf{\codefont forall}}\;\Varid{a}\;\Varid{b}\;\Varid{c}\;(\Varid{f}\mathbin{:}\Varid{a}\mathbin{\text{\codefont ->}}\Varid{b})\;(\Varid{g}\mathbin{:}\Varid{b}\mathbin{\text{\codefont ->}}\Varid{c})\;(\Varid{x}\mathbin{:}\Varid{t}\;\Varid{a}),{}\<[E]%
\\[-0.3ex]%
\>[9]{}\hsindent{3}{}\<[12]%
\>[12]{}\Varid{fmap}\;\Varid{g}\;(\Varid{fmap}\;\Varid{f}\;\Varid{x})\mathrel{=}\Varid{fmap}\;(\Varid{g}\mathbin{∘}\Varid{f})\;\Varid{x}\mathord{\text{\codefont \}}}\mathop{\text{.}}{}\<[E]%
\ColumnHook
\end{hscode}\resethooks
A \ensuremath{\Varid{list}} instance of the \ensuremath{\Conid{FunctorLaws}} type class is a formal proof that the
\ensuremath{\Varid{list}} type, using this definition of \ensuremath{\Varid{map}}, is a lawful
functor.

This process makes sense only for \emph{inductive} data types and
\emph{total, terminating} functions. This is where the semantics of lazy and strict evaluation, and hence of Haskell and Coq, coincide~\cite{fast-and-loose}.
However, the payoff is that a successful translation is itself a termination
proof, even before other properties have been
shown. Furthermore, because Coq programs may be evaluated (within Coq) or
compiled (via extraction) these properties apply, not to a formal model of
computation, but to actual runnable code.

Our overarching goal is to make it easy for a Haskell programmer to produce Coq
versions of their programs that are suitable for verification. The Coq
rendition should closely follow the Haskell code -- the same names should be used,
even within functions; types should be unaltered; abstractions like type classes and
modules should be preserved -- so that the programmer obtains not just a
black-box that happens to do the same thing as the original Haskell program,
but a \emph{live} Coq version of the input.

Furthermore, the development environment should include as much as possible of
\emph{total} Haskell. In particular, programmers should have access to
standard libraries and language features and face few limitations other than
totality. \jp{?}Also, because programs often change, the
generated Coq must be usable directly, or with \emph{declarative}
modifications, so that the proofs can evolve with the program.

\jp{I'm not totally convinced that the tools are so much better for prototyping
  in Haskell than Coq, esp.\ now that we have quickchick!  Also since you would
  have to write in a fragment of Haskell}Conversely, an additional application
of \texttt{hs-to-coq} is as a Haskell ``rapid prototyping front-end'' for Coq. A
potential workflow is: (1) implement a program in Haskell first, in order to
quickly develop and test it; (2) use \texttt{hs-to-coq} to translate it to the
Coq world; and (3) extend and verify the Coq output. This framework allows
diverse groups of functional programmers and proof engineers to collaborate;
focusing on their areas of expertise.

Therefore, in this paper, we describe the design and implementation of the
\texttt{hs-to-coq} tool and our experiences with its application in several
domains.  In particular, the contributions of this paper are as follows.
\begin{itemize}
\item We describe the use of our methodology and tool in three different
  examples, showing how it can be used to state and prove the monad laws,
  replicate textbook equational reasoning, and verify data structure
  invariants (\cref{sec:case-studies}).
\item We identify several design considerations in the development of the
  \texttt{hs-to-coq} tool itself and discuss our approach to resolving the
  differences between Haskell and Coq (\cref{sec:design}).
\item We discuss a Coq translation of the Haskell \text{\tt base} library for working
  with translated programs that we have developed using \texttt{hs-to-coq}
  (\cref{sec:base}).
\end{itemize}
We discuss related work in \cref{sec:related} and future directions in
\cref{sec:conclusion}. Our tool, \text{\tt base} libraries, and the case studies are
freely available as open source
software.\footnote{\url{https://github.com/antalsz/hs-to-coq}}\jp{intro is a
  little hard to follow}

\section{Reasoning About Haskell Code in Coq}
\label{sec:case-studies}

\asz{General note: need to do a consistency pass for tense}%
We present and evaluate our approach to verifying Haskell in three examples,
all involving pre-existing Haskell code.%
\iflong
\scw{We can cut this list of bullets. It doesn't add much.}
\begin{itemize}
\item In \cref{sec:successors}, we further investigate how our method applies to
  type classes and their laws.
\item In \cref{sec:compiler}, we apply our method to a textbook example, and
  consider how we may need to ensure that Haskell code is amenable to
  verification.\asz{Ugh, wordy, revisit.}
\item In \cref{sec:bags}, we demonstrate that we can apply our methodology to
  portions of code taken out of a larger project.\asz{Mention edits?}
\end{itemize}
\fi

%
%

\subsection{Algebraic Laws}\label{sec:successors}

\newcommand{\figsuccs}{
\begin{figure}
\abovedisplayskip=0pt
\belowdisplayskip=0pt
\raggedright
\begin{hscode}\SaveRestoreHook
\column{B}{@{}>{\hspre}l<{\hspost}@{}}%
\column{5}{@{}>{\hspre}l<{\hspost}@{}}%
\column{7}{@{}>{\hspre}l<{\hspost}@{}}%
\column{9}{@{}>{\hspre}l<{\hspost}@{}}%
\column{E}{@{}>{\hspre}l<{\hspost}@{}}%
\>[B]{}\textbf{\codefont module}\;\Conid{\Conid{Control}.\Conid{Applicative}.Successors}\;\textbf{\codefont where}{}\<[E]%
\\[\blanklineskip]%
\>[B]{}\textbf{\codefont data}\;\Conid{Succs}\;\Varid{a}\mathrel{=}\Conid{Succs}\;\Varid{a}\;[\mskip1.5mu \Varid{a}\mskip1.5mu]\;\textbf{\codefont deriving}\;(\Conid{Show},\Conid{Eq}){}\<[E]%
\\[\blanklineskip]%
\>[B]{}\Varid{getCurrent}\mathbin{::}\Conid{Succs}\;\Varid{t}\mathbin{\text{\codefont ->}}\Varid{t}{}\<[E]%
\\[-0.3ex]%
\>[B]{}\Varid{getCurrent}\;(\Conid{Succs}\;\Varid{x}\;\anonymous )\mathrel{=}\Varid{x}{}\<[E]%
\\[\blanklineskip]%
\>[B]{}\Varid{getSuccs}\mathbin{::}\Conid{Succs}\;\Varid{t}\mathbin{\text{\codefont ->}}[\mskip1.5mu \Varid{t}\mskip1.5mu]{}\<[E]%
\\[-0.3ex]%
\>[B]{}\Varid{getSuccs}\;(\Conid{Succs}\;\anonymous \;\Varid{xs})\mathrel{=}\Varid{xs}{}\<[E]%
\\[\blanklineskip]%
\>[B]{}\textbf{\codefont instance}\;\Conid{Functor}\;\Conid{Succs}\;\textbf{\codefont where}{}\<[E]%
\\[-0.3ex]%
\>[B]{}\hsindent{5}{}\<[5]%
\>[5]{}\Varid{fmap}\;\Varid{f}\;(\Conid{Succs}\;\Varid{x}\;\Varid{xs})\mathrel{=}\Conid{Succs}\;(\Varid{f}\;\Varid{x})\;(\Varid{map}\;\Varid{f}\;\Varid{xs}){}\<[E]%
\\[\blanklineskip]%
\>[B]{}\textbf{\codefont instance}\;\Conid{Applicative}\;\Conid{Succs}\;\textbf{\codefont where}{}\<[E]%
\\[-0.3ex]%
\>[B]{}\hsindent{5}{}\<[5]%
\>[5]{}\Varid{pure}\;\Varid{x}\mathrel{=}\Conid{Succs}\;\Varid{x}\;[\mskip1.5mu \mskip1.5mu]{}\<[E]%
\\[-0.3ex]%
\>[B]{}\hsindent{5}{}\<[5]%
\>[5]{}\Conid{Succs}\;\Varid{f}\;\Varid{fs}\mathbin{\text{\codefont <*>}}\Conid{Succs}\;\Varid{x}\;\Varid{xs}{}\<[E]%
\\[-0.3ex]%
\>[5]{}\hsindent{4}{}\<[9]%
\>[9]{}\mathrel{=}\Conid{Succs}\;(\Varid{f}\;\Varid{x})\;(\Varid{map}\;(\text{\codefont \$}\Varid{x})\;\Varid{fs}\mathbin{\text{\codefont ++}}\Varid{map}\;\Varid{f}\;\Varid{xs}){}\<[E]%
\\[\blanklineskip]%
\>[B]{}\textbf{\codefont instance}\;\Conid{Monad}\;\Conid{Succs}\;\textbf{\codefont where}{}\<[E]%
\\[-0.3ex]%
\>[B]{}\hsindent{5}{}\<[5]%
\>[5]{}\Conid{Succs}\;\Varid{x}\;\Varid{xs}\mathbin{\text{\codefont >>=}}\Varid{f}{}\<[E]%
\\[-0.3ex]%
\>[5]{}\hsindent{4}{}\<[9]%
\>[9]{}\mathrel{=}\Conid{Succs}\;\Varid{y}\;(\Varid{map}\;(\Varid{getCurrent}\mathop{\text{.}}\Varid{f})\;\Varid{xs}\mathbin{\text{\codefont ++}}\Varid{ys}){}\<[E]%
\\[-0.3ex]%
\>[5]{}\hsindent{2}{}\<[7]%
\>[7]{}\textbf{\codefont where}\;\Conid{Succs}\;\Varid{y}\;\Varid{ys}\mathrel{=}\Varid{f}\;\Varid{x}{}\<[E]%
\ColumnHook
\end{hscode}\resethooks
\caption{The \text{\tt successors} library}
\label{fig:succs}
\end{figure}
}

\newif\ifFigMonadLawsNarrow
\FigMonadLawsNarrowtrue

\newcommand*{\FigMonadLawsEnvironment}{%
  \ifFigMonadLawsNarrow
    figure%
  \else
    figure*%
  \fi
}

\newcommand*{\FigMonadLawsNarrow}{%
  \includegraphics[%
    width=\linewidth,%
    page=1,%
    clip,%
    trim={1.7cm 3.3cm 1.7cm 40.14cm}%
  ]{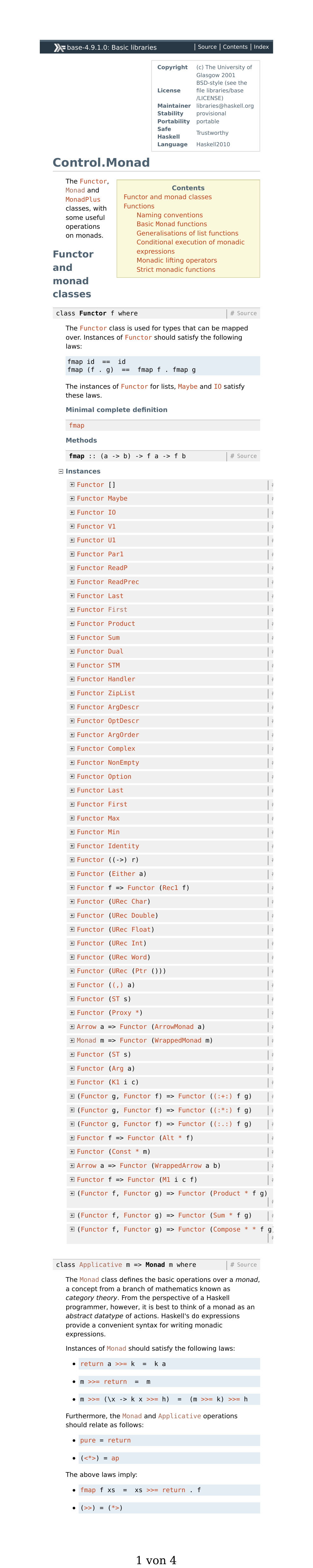}%
}

\newcommand*{\FigMonadLawsWide}{%
  \includegraphics[%
    width=\linewidth,%
    page=2,%
    clip,%
    trim={2cm 3.5cm 2cm 18.8cm}%
  ]{monad-docs.pdf}%
}

\newcommand*{\FigMonadLawsCaption}{%
  The documentation of the \ensuremath{\Conid{Monad}} type class lists the three monad
  laws and the two laws relating it to \ensuremath{\Conid{Applicative}} (screenshot).%
}

\newcommand*{\FigMonadLaws}{%
  \begin{\FigMonadLawsEnvironment}
    \ifFigMonadLawsNarrow
      \FigMonadLawsNarrow
    \else
      \FigMonadLawsWide
    \fi
    \caption[\FigMonadLawsCaption]{\FigMonadLawsCaption\footnotemark}
    \label{fig:monad-laws-comment}
  \end{\FigMonadLawsEnvironment}%
}

\FigMonadLaws
\footnotetext{%
  \url{http://hackage.haskell.org/package/base-4.9.1.0/docs/Prelude.html\#t:Monad}}

\paragraph*{Objective}
The \ensuremath{\Conid{Functor}} type class is not the only class with laws. Many Haskell
programs feature structures that are not only instances of the \ensuremath{\Conid{Functor}} class,
but also of \ensuremath{\Conid{Applicative}} and \ensuremath{\Conid{Monad}} as well.  All three of these classes come
with laws. Library authors are expected to establish that their instances of these
classes are lawful (respect the laws). Programmers using their libraries may then
use these laws to reason about their code.

\asz{Reviewer 1 says: ``It would be worth showing the types of the monad
  operations'', and similarly for \ensuremath{\Varid{ap}}.  But that doesn't show up until after a
  lot more documentation we don't care about, and looks really ugly in the
  narrow format.}%
\scw{I agree that we shouldn't do this. We do eventually show the (Coq) type
  for return/bind. We never discuss ap, but I don't think we need to.}%
\scw{What is the \cn in the caption?}
For example, the documentation for the \ensuremath{\Conid{Monad}} type class, shown in
\cref{fig:monad-laws-comment}, lists the three standard \ensuremath{\Conid{Monad}} laws as well as
two more laws that connect the \ensuremath{\Conid{Monad}} methods to those of its superclass
\ensuremath{\Conid{Applicative}}.  Typically, reasoning about these laws is done on paper, but our
tool makes mechanical verification available.

\sloppy

In this first example, we take the open source \text{\tt successors}
library~\cite{successors-0.1}
and show that its instances of the \ensuremath{\Conid{Functor}}, \ensuremath{\Conid{Applicative}}, and \ensuremath{\Conid{Monad}} classes
are lawful.  This library provides a type \ensuremath{\Conid{Succs}} that represents one step in a
nondeterministic reduction relation; the type class instances allow us to
combine two relations into one that takes a single step from either of the
original relations. \Cref{fig:succs} shows the complete, unmodified code of the
library. The source code also contains, as a comment,~80 lines of manual
equational reasoning establishing the type class laws.

\fussy

\figsuccs

\newcommand{\figsuccsv}{
\begin{figure}
\abovedisplayskip=0pt
\belowdisplayskip=0pt
\raggedright
\begin{hscode}\SaveRestoreHook
\column{B}{@{}>{\hspre}l<{\hspost}@{}}%
\column{3}{@{}>{\hspre}l<{\hspost}@{}}%
\column{5}{@{}>{\hspre}l<{\hspost}@{}}%
\column{7}{@{}>{\hspre}l<{\hspost}@{}}%
\column{9}{@{}>{\hspre}l<{\hspost}@{}}%
\column{11}{@{}>{\hspre}l<{\hspost}@{}}%
\column{E}{@{}>{\hspre}l<{\hspost}@{}}%
\>[B]{}{\textbf{\codefont Inductive}}\;\Conid{Succs}\;\Varid{a}\mathbin{:}\Conid{Type}\mathbin{:=}{}\<[E]%
\\[-0.3ex]%
\>[B]{}\hsindent{3}{}\<[3]%
\>[3]{}\Conid{Mk\char95 Succs}\mathbin{:}\Varid{a}\mathbin{\text{\codefont ->}}\Varid{list}\;\Varid{a}\mathbin{\text{\codefont ->}}\Conid{Succs}\;\Varid{a}\mathop{\text{.}}{}\<[E]%
\\[\blanklineskip]%
\>[B]{}(\mathbin{*} \textit{~Instances for Functor and Applicative omitted.~} \mathbin{*}){}\<[E]%
\\[\blanklineskip]%
\>[B]{}{\textbf{\codefont Local}}\;{\textbf{\codefont Definition}}\;\Varid{instance\char95 Monad\char95 Succs\char95 op\char95 zgzgze\char95 \char95 }{}\<[E]%
\\[-0.3ex]%
\>[B]{}\hsindent{3}{}\<[3]%
\>[3]{}\mathbin{:}{\textbf{\codefont forall}}\;\mathord{\text{\codefont \{}}\Varid{a}\mathord{\text{\codefont \}}}\;\mathord{\text{\codefont \{}}\Varid{b}\mathord{\text{\codefont \}}},\Conid{Succs}\;\Varid{a}\mathbin{\text{\codefont ->}}(\Varid{a}\mathbin{\text{\codefont ->}}\Conid{Succs}\;\Varid{b})\mathbin{\text{\codefont ->}}\Conid{Succs}\;\Varid{b}{}\<[E]%
\\[-0.3ex]%
\>[B]{}\hsindent{3}{}\<[3]%
\>[3]{}\mathbin{:=}{}\<[7]%
\>[7]{}{\textbf{\codefont fun}}\;\mathord{\text{\codefont \{}}\Varid{a}\mathord{\text{\codefont \}}}\;\mathord{\text{\codefont \{}}\Varid{b}\mathord{\text{\codefont \}}}\mathbin{\text{\codefont =>}}{\textbf{\codefont fun}}\;\Varid{arg\char95 4\char95 \char95 }\;\Varid{arg\char95 5\char95 \char95 }\mathbin{\text{\codefont =>}}{}\<[E]%
\\[-0.3ex]%
\>[7]{}{\textbf{\codefont match}}\;\Varid{arg\char95 4\char95 \char95 },\Varid{arg\char95 5\char95 \char95 }\;{\textbf{\codefont with}}{}\<[E]%
\\[-0.3ex]%
\>[7]{}\mid \Conid{Mk\char95 Succs}\;\Varid{x}\;\Varid{xs},\Varid{f}\mathbin{\text{\codefont =>}}{\textbf{\codefont match}}\;\Varid{f}\;\Varid{x}\;{\textbf{\codefont with}}{}\<[E]%
\\[-0.3ex]%
\>[7]{}\hsindent{2}{}\<[9]%
\>[9]{}\mid \Conid{Mk\char95 Succs}\;\Varid{y}\;\Varid{ys}\mathbin{\text{\codefont =>}}\Conid{Mk\char95 Succs}\;\Varid{y}{}\<[E]%
\\[-0.3ex]%
\>[9]{}\hsindent{2}{}\<[11]%
\>[11]{}(\Varid{app}\;(\Varid{map}\;(\Varid{compose}\;\Varid{getCurrent}\;\Varid{f})\;\Varid{xs})\;\Varid{ys}){}\<[E]%
\\[-0.3ex]%
\>[7]{}\hsindent{2}{}\<[9]%
\>[9]{}{\textbf{\codefont end}}{}\<[E]%
\\[-0.3ex]%
\>[7]{}{\textbf{\codefont end}}\mathop{\text{.}}{}\<[E]%
\\[\blanklineskip]%
\>[B]{}{\textbf{\codefont Local}}\;{\textbf{\codefont Definition}}\;\Varid{instance\char95 Monad\char95 Succs\char95 return\char95 }{}\<[E]%
\\[-0.3ex]%
\>[B]{}\hsindent{3}{}\<[3]%
\>[3]{}\mathbin{:}{\textbf{\codefont forall}}\;\mathord{\text{\codefont \{}}\Varid{a}\mathord{\text{\codefont \}}},\Varid{a}\mathbin{\text{\codefont ->}}\Conid{Succs}\;\Varid{a}\mathbin{:=}{\textbf{\codefont fun}}\;\mathord{\text{\codefont \{}}\Varid{a}\mathord{\text{\codefont \}}}\mathbin{\text{\codefont =>}}\Varid{pure}\mathop{\text{.}}{}\<[E]%
\\[\blanklineskip]%
\>[B]{}{\textbf{\codefont Local}}\;{\textbf{\codefont Definition}}\;\Varid{instance\char95 Monad\char95 Succs\char95 op\char95 zgzg\char95 \char95 }{}\<[E]%
\\[-0.3ex]%
\>[B]{}\hsindent{3}{}\<[3]%
\>[3]{}\mathbin{:}{\textbf{\codefont forall}}\;\mathord{\text{\codefont \{}}\Varid{a}\mathord{\text{\codefont \}}}\;\mathord{\text{\codefont \{}}\Varid{b}\mathord{\text{\codefont \}}},\Conid{Succs}\;\Varid{a}\mathbin{\text{\codefont ->}}\Conid{Succs}\;\Varid{b}\mathbin{\text{\codefont ->}}\Conid{Succs}\;\Varid{b}{}\<[E]%
\\[-0.3ex]%
\>[B]{}\hsindent{3}{}\<[3]%
\>[3]{}\mathbin{:=}{\textbf{\codefont fun}}\;\mathord{\text{\codefont \{}}\Varid{a}\mathord{\text{\codefont \}}}\;\mathord{\text{\codefont \{}}\Varid{b}\mathord{\text{\codefont \}}}\mathbin{\text{\codefont =>}}\Varid{op\char95 ztzg\char95 \char95 }\mathop{\text{.}}{}\<[E]%
\\[\blanklineskip]%
\>[B]{}{\textbf{\codefont Instance}}\;\Varid{instance\char95 Monad\char95 Succs}\mathbin{:}\Conid{Monad}\;\Conid{Succs}\mathbin{:=}\mathord{\text{\codefont \{}}{}\<[E]%
\\[-0.3ex]%
\>[B]{}\hsindent{3}{}\<[3]%
\>[3]{}\Varid{op\char95 zgzg\char95 \char95 }\mathbin{:=}{\textbf{\codefont fun}}\;\mathord{\text{\codefont \{}}\Varid{a}\mathord{\text{\codefont \}}}\;\mathord{\text{\codefont \{}}\Varid{b}\mathord{\text{\codefont \}}}\mathbin{\text{\codefont =>}}{}\<[E]%
\\[-0.3ex]%
\>[3]{}\hsindent{2}{}\<[5]%
\>[5]{}\Varid{instance\char95 Monad\char95 Succs\char95 op\char95 zgzg\char95 \char95 };{}\<[E]%
\\[-0.3ex]%
\>[B]{}\hsindent{3}{}\<[3]%
\>[3]{}\Varid{op\char95 zgzgze\char95 \char95 }\mathbin{:=}{\textbf{\codefont fun}}\;\mathord{\text{\codefont \{}}\Varid{a}\mathord{\text{\codefont \}}}\;\mathord{\text{\codefont \{}}\Varid{b}\mathord{\text{\codefont \}}}\mathbin{\text{\codefont =>}}{}\<[E]%
\\[-0.3ex]%
\>[3]{}\hsindent{2}{}\<[5]%
\>[5]{}\Varid{instance\char95 Monad\char95 Succs\char95 op\char95 zgzgze\char95 \char95 };{}\<[E]%
\\[-0.3ex]%
\>[B]{}\hsindent{3}{}\<[3]%
\>[3]{}\Varid{return\char95 }\mathbin{:=}{\textbf{\codefont fun}}\;\mathord{\text{\codefont \{}}\Varid{a}\mathord{\text{\codefont \}}}\mathbin{\text{\codefont =>}}{}\<[E]%
\\[-0.3ex]%
\>[3]{}\hsindent{2}{}\<[5]%
\>[5]{}\Varid{instance\char95 Monad\char95 Succs\char95 return\char95 }\mathord{\text{\codefont \}}}\mathop{\text{.}}{}\<[E]%
\ColumnHook
\end{hscode}\resethooks
\caption{Excerpt of the Coq code produced from \cref{fig:succs}.
  (To fit the available width, module prefixes are omitted and lines are
  manually re-wrapped.)}
\label{fig:succs.v}
\end{figure}}

\paragraph*{Experience}
\Cref{fig:succs.v} shows the generated Coq code for the type \ensuremath{\Conid{Succs}} and the
\ensuremath{\Conid{Monad}} instance. The first line is the corresponding definition of the \ensuremath{\Conid{Succs}}
data type. Because the Haskell program uses the same name for both the type
constructor \ensuremath{\Conid{Succs}} and its single data constructor, \texttt{hs-to-coq}
automatically renames the latter to \ensuremath{\Conid{Mk\char95 Succs}} to avoid this name
conflict.\footnotemark

The rest of the figure contains the instance of the \ensuremath{\Conid{Monad}} type class for the
\ensuremath{\Conid{Succs}} type.  This code imports a Coq version of Haskell's standard library
\text{\tt base} that we have also developed using \texttt{hs-to-coq} (see
\cref{sec:base}). The \ensuremath{\Conid{Monad}} type class from that library, shown below, is a
direct translation of GHC's implementation of the base libraries.

\footnotetext{The prefix \ensuremath{\Conid{Mk\char95 }} is almost never used for Haskell names, so this
  heuristic is very unlikely to produce a constructor name that clashes with an
  existing Haskell name.  The \text{\tt hs\char45{}to\char45{}coq} tool includes the ability to customize
  renaming, which can be used in case there are name clashes; see
  \cref{sec:what-is-in-base} for more details.}

\label{class-monad}%
\begin{hscode}\SaveRestoreHook
\column{B}{@{}>{\hspre}l<{\hspost}@{}}%
\column{3}{@{}>{\hspre}l<{\hspost}@{}}%
\column{E}{@{}>{\hspre}l<{\hspost}@{}}%
\>[B]{}{\textbf{\codefont Class}}\;\Conid{Monad}\;\Varid{m}\text{\phantom{x}}\text{\codefont\textasciigrave}\mathord{\text{\codefont \{}}\Conid{Applicative}\;\Varid{m}\mathord{\text{\codefont \}}}\mathbin{:=}\mathord{\text{\codefont \{}}{}\<[E]%
\\[-0.3ex]%
\>[B]{}\hsindent{3}{}\<[3]%
\>[3]{}\Varid{op\char95 zgzg\char95 \char95 }\mathbin{:}{\textbf{\codefont forall}}\;\mathord{\text{\codefont \{}}\Varid{a}\mathord{\text{\codefont \}}}\;\mathord{\text{\codefont \{}}\Varid{b}\mathord{\text{\codefont \}}},\Varid{m}\;\Varid{a}\mathbin{\text{\codefont ->}}\Varid{m}\;\Varid{b}\mathbin{\text{\codefont ->}}\Varid{m}\;\Varid{b};{}\<[E]%
\\[-0.3ex]%
\>[B]{}\hsindent{3}{}\<[3]%
\>[3]{}\Varid{op\char95 zgzgze\char95 \char95 }\mathbin{:}{\textbf{\codefont forall}}\;\mathord{\text{\codefont \{}}\Varid{a}\mathord{\text{\codefont \}}}\;\mathord{\text{\codefont \{}}\Varid{b}\mathord{\text{\codefont \}}},\Varid{m}\;\Varid{a}\mathbin{\text{\codefont ->}}(\Varid{a}\mathbin{\text{\codefont ->}}\Varid{m}\;\Varid{b})\mathbin{\text{\codefont ->}}\Varid{m}\;\Varid{b};{}\<[E]%
\\[-0.3ex]%
\>[B]{}\hsindent{3}{}\<[3]%
\>[3]{}\Varid{return\char95 }\mathbin{:}{\textbf{\codefont forall}}\;\mathord{\text{\codefont \{}}\Varid{a}\mathord{\text{\codefont \}}},\Varid{a}\mathbin{\text{\codefont ->}}\Varid{m}\;\Varid{a}\mathord{\text{\codefont \}}}\mathop{\text{.}}{}\<[E]%
\\[\blanklineskip]%
\>[B]{}{\textbf{\codefont Infix}}\;\text{\codefont\textquotedbl}\mathbin{\text{\codefont >>}}\text{\codefont\textquotedbl}\mathbin{:=}(\Varid{op\char95 zgzg\char95 \char95 })\;(\Varid{at}\;\Varid{level}\;\mathrm{99})\mathop{\text{.}}{}\<[E]%
\\[-0.3ex]%
\>[B]{}{\textbf{\codefont Notation}}\;\text{\codefont\textquotedbl\textquotesingle\_}\mathbin{\text{\codefont >>}}\text{\codefont\_\textquotesingle\textquotedbl}\mathbin{:=}(\Varid{op\char95 zgzg\char95 \char95 })\mathop{\text{.}}{}\<[E]%
\\[-0.3ex]%
\>[B]{}{\textbf{\codefont Infix}}\;\text{\codefont\textquotedbl}\mathbin{\text{\codefont >>=}}\text{\codefont\textquotedbl}\mathbin{:=}(\Varid{op\char95 zgzgze\char95 \char95 })\;(\Varid{at}\;\Varid{level}\;\mathrm{99})\mathop{\text{.}}{}\<[E]%
\\[-0.3ex]%
\>[B]{}{\textbf{\codefont Notation}}\;\text{\codefont\textquotedbl\textquotesingle\_}\mathbin{\text{\codefont >>=}}\text{\codefont\_\textquotesingle\textquotedbl}\mathbin{:=}(\Varid{op\char95 zgzgze\char95 \char95 })\mathop{\text{.}}{}\<[E]%
\ColumnHook
\end{hscode}\resethooks

\figsuccsv

As in Haskell, the \ensuremath{\Conid{Monad}} class includes the \ensuremath{\Varid{return}} and \ensuremath{\mathbin{\text{\codefont >>=}}} methods, which
form the mathematical definition of a monad, as well as an additional sequencing
method \ensuremath{\mathbin{\text{\codefont >>}}}. Again due to restrictions on naming, the Coq version uses
alternative names for all three of these methods. As \ensuremath{\Varid{return}} is a keyword, it
is replaced with \ensuremath{\Varid{return\char95 }}. Furthermore, Coq does not support variables with
symbolic names, so the bind and sequencing operators are replaced by names
starting with \ensuremath{\Varid{op\char95 }}.  In \cref{fig:succs.v}, we can see: \ensuremath{\Varid{op\char95 zgzgze\char95 \char95 }}, the
translation of \ensuremath{\mathbin{\text{\codefont >>=}}}; \ensuremath{\Varid{op\char95 zgzg\char95 \char95 }}, the translation of \ensuremath{\mathbin{\text{\codefont >>}}}; and \ensuremath{\Varid{op\char95 ztzg\char95 \char95 }}, the
translation of \ensuremath{\mathbin{\text{\codefont *>}}} from the \ensuremath{\Conid{Applicative}} type class.  These names are
systematically derived using GHC’s ``Z-encoding''.  Haskell's \ensuremath{\mathbin{\text{\codefont ++}}} operator is
translated to the pre-existing Coq \ensuremath{\Varid{app}} function, so it does not receive an
\ensuremath{\Varid{op\char95 }} name.

Note that our version of the \ensuremath{\Conid{Monad}} type class does not include the
infamous method \ensuremath{\Varid{fail}\mathbin{::}\Conid{Monad}\;\Varid{m}\mathbin{\text{\codefont =>}}\Conid{String}\mathbin{\text{\codefont ->}}\Varid{m}\;\Varid{a}}. For many monads, including
\ensuremath{\Conid{Succs}}, a function with this type signature is impossible to implement in Coq
-- this method is frequently partial.\footnote{In fact, this is considered to be
  a problem in Haskell as well, so the method is currently being moved into its
  own class, \ensuremath{\Conid{MonadFail}}; we translate this class (in the module
  \ensuremath{\Conid{\Conid{Control}.\Conid{Monad}.Fail}}) as well, for monads that have total definitions of
  this operation.} As a result, we have instructed \texttt{hs-to-coq} to skip
this method when translating the \ensuremath{\Conid{Monad}} class and its instances.

The instance of the \ensuremath{\Conid{Monad}} class in \cref{fig:succs.v} includes definitions for
all three members of the class. The first definition is translated from the \ensuremath{\mathbin{\text{\codefont >>=}}} method of
the input file; \texttt{hs-to-coq} supplies the other two
components from the default definitions in the \ensuremath{\Conid{Monad}} class.

\newcommand{\figmonadlawcoq}{
\begin{figure}
\abovedisplayskip=0pt
\belowdisplayskip=0pt
\begin{hscode}\SaveRestoreHook
\column{B}{@{}>{\hspre}l<{\hspost}@{}}%
\column{3}{@{}>{\hspre}l<{\hspost}@{}}%
\column{5}{@{}>{\hspre}l<{\hspost}@{}}%
\column{7}{@{}>{\hspre}l<{\hspost}@{}}%
\column{8}{@{}>{\hspre}l<{\hspost}@{}}%
\column{25}{@{}>{\hspre}c<{\hspost}@{}}%
\column{25E}{@{}l@{}}%
\column{27}{@{}>{\hspre}c<{\hspost}@{}}%
\column{27E}{@{}l@{}}%
\column{28}{@{}>{\hspre}l<{\hspost}@{}}%
\column{30}{@{}>{\hspre}l<{\hspost}@{}}%
\column{38}{@{}>{\hspre}c<{\hspost}@{}}%
\column{38E}{@{}l@{}}%
\column{41}{@{}>{\hspre}l<{\hspost}@{}}%
\column{E}{@{}>{\hspre}l<{\hspost}@{}}%
\>[B]{}{\textbf{\codefont Class}}\;\Conid{MonadLaws}\;(\Varid{t}\mathbin{:}\Conid{Type}\mathbin{\text{\codefont ->}}\Conid{Type}){}\<[E]%
\\[-0.3ex]%
\>[B]{}\hsindent{5}{}\<[5]%
\>[5]{}\text{\codefont\textasciigrave}\mathord{\text{\codefont \{}}\mathord{\text{\codefont !}} \Conid{Functor}\;\Varid{t},\mathord{\text{\codefont !}} \Conid{Applicative}\;\Varid{t},\mathord{\text{\codefont !}} \Conid{Monad}\;\Varid{t},{}\<[E]%
\\[-0.3ex]%
\>[5]{}\hsindent{2}{}\<[7]%
\>[7]{}\mathord{\text{\codefont !}} \Conid{FunctorLaws}\;\Varid{t},\mathord{\text{\codefont !}} \Conid{ApplicativeLaws}\;\Varid{t}\mathord{\text{\codefont \}}}\mathbin{:=}{}\<[E]%
\\[-0.3ex]%
\>[B]{}\hsindent{3}{}\<[3]%
\>[3]{}\mathord{\text{\codefont \{}}\Varid{monad\char95 left\char95 id}\mathbin{:}{\textbf{\codefont forall}}\;\Conid{A}\;\Conid{B}\;(\Varid{a}\mathbin{:}\Conid{A})\;(\Varid{k}\mathbin{:}\Conid{A}\mathbin{\text{\codefont ->}}\Varid{t}\;\Conid{B}),{}\<[E]%
\\[-0.3ex]%
\>[3]{}\hsindent{5}{}\<[8]%
\>[8]{}(\Varid{return\char95 }\;\Varid{a}\mathbin{\text{\codefont >>=}}\Varid{k}){}\<[27]%
\>[27]{}\mathrel{=}{}\<[27E]%
\>[30]{}(\Varid{k}\;\Varid{a});{}\<[E]%
\\[-0.3ex]%
\>[3]{}\hsindent{2}{}\<[5]%
\>[5]{}\Varid{monad\char95 right\char95 id}\mathbin{:}{\textbf{\codefont forall}}\;\Conid{A}\;(\Varid{m}\mathbin{:}\Varid{t}\;\Conid{A}),{}\<[E]%
\\[-0.3ex]%
\>[5]{}\hsindent{3}{}\<[8]%
\>[8]{}(\Varid{m}\mathbin{\text{\codefont >>=}}\Varid{return\char95 }){}\<[25]%
\>[25]{}\mathrel{=}{}\<[25E]%
\>[28]{}\Varid{m};{}\<[E]%
\\[-0.3ex]%
\>[3]{}\hsindent{2}{}\<[5]%
\>[5]{}\Varid{monad\char95 composition}\mathbin{:}{\textbf{\codefont forall}}\;\Conid{A}\;\Conid{B}\;\Conid{C}{}\<[E]%
\\[-0.3ex]%
\>[5]{}\hsindent{3}{}\<[8]%
\>[8]{}(\Varid{m}\mathbin{:}\Varid{t}\;\Conid{A})\;(\Varid{k}\mathbin{:}\Conid{A}\mathbin{\text{\codefont ->}}\Varid{t}\;\Conid{B})\;(\Varid{h}\mathbin{:}\Conid{B}\mathbin{\text{\codefont ->}}\Varid{t}\;\Conid{C}),{}\<[E]%
\\[-0.3ex]%
\>[5]{}\hsindent{3}{}\<[8]%
\>[8]{}(\Varid{m}\mathbin{\text{\codefont >>=}}({\textbf{\codefont fun}}\;\Varid{x}\mathbin{\text{\codefont =>}}\Varid{k}\;\Varid{x}\mathbin{\text{\codefont >>=}}\Varid{h})){}\<[38]%
\>[38]{}\mathrel{=}{}\<[38E]%
\>[41]{}((\Varid{m}\mathbin{\text{\codefont >>=}}\Varid{k})\mathbin{\text{\codefont >>=}}\Varid{h});{}\<[E]%
\\[-0.3ex]%
\>[3]{}\hsindent{2}{}\<[5]%
\>[5]{}\Varid{monad\char95 applicative\char95 pure}\mathbin{:}{\textbf{\codefont forall}}\;\Conid{A}\;(\Varid{x}\mathbin{:}\Conid{A}),{}\<[E]%
\\[-0.3ex]%
\>[5]{}\hsindent{3}{}\<[8]%
\>[8]{}\Varid{pure}\;\Varid{x}\mathrel{=}\Varid{return\char95 }\;\Varid{x};{}\<[E]%
\\[-0.3ex]%
\>[3]{}\hsindent{2}{}\<[5]%
\>[5]{}\Varid{monad\char95 applicative\char95 ap}\mathbin{:}{\textbf{\codefont forall}}\;\Conid{A}\;\Conid{B}{}\<[E]%
\\[-0.3ex]%
\>[5]{}\hsindent{3}{}\<[8]%
\>[8]{}(\Varid{f}\mathbin{:}\Varid{t}\;(\Conid{A}\mathbin{\text{\codefont ->}}\Conid{B}))\;(\Varid{x}\mathbin{:}\Varid{t}\;\Conid{A}),{}\<[E]%
\\[-0.3ex]%
\>[5]{}\hsindent{3}{}\<[8]%
\>[8]{}(\Varid{f}\mathbin{\text{\codefont <*>}}\Varid{x})\mathrel{=}\Varid{ap}\;\Varid{f}\;\Varid{x}\mathord{\text{\codefont \}}}\mathop{\text{.}}{}\<[E]%
\ColumnHook
\end{hscode}\resethooks
\caption{Coq type class capturing the \ensuremath{\Conid{Monad}} laws.}
\label{fig:monad-laws-coq}
\end{figure}}

\sloppy

Our \text{\tt base} library also includes an additional type class formalizing the laws for the \ensuremath{\Conid{Monad}}
class, shown in \cref{fig:monad-laws-coq}.\jp{this is not obtained using
  \texttt{hs-to-coq}} These laws directly correspond to the documentation in
\cref{fig:monad-laws-comment}. Using this definition (and similar ones for
\ensuremath{\Conid{FunctorLaws}} and \ensuremath{\Conid{ApplicativeLaws}}), we can show that the Coq implementation
satisfies the requirements of this class. These proofs are
straightforward and are analogous to the reasoning found in the handwritten~80 line
comment in the library.

\fussy

\figmonadlawcoq

\paragraph*{Conclusion}
The proofs about \ensuremath{\Conid{Succs}} demonstrate that we can translate Haskell code that
uses type classes and instances using Coq’s support for type classes. We can
then use Coq to perform reasoning that was previously done manually, and we
can support this further by capturing the requirements of type classes in
additional type classes.

\subsection{Hutton's Razor}\label{sec:compiler}

\paragraph*{Objective}

Our next case study is ``Hutton's razor'', from \emph{Programming in
  Haskell}~\cite{Hutton:2016:PH:3092752}.  It includes a small expression
language with an interpreter and a simple compiler from this language to a stack
machine~\cite[Section~16.7]{Hutton:2016:PH:3092752}.  We present our version of
his code in \cref{fig:compiler}.

Hutton uses this example to demonstrate how equational reasoning can be used to
show compiler correctness.  In other words, Hutton shows that executing the
output of the compiler with an empty stack produces the same result as
evaluating an expression:
\[
  \ensuremath{\Varid{exec}\;(\Varid{comp}\;\Varid{e})\;[\mskip1.5mu \mskip1.5mu]} = \ensuremath{\Conid{Just}\;[\mskip1.5mu \Varid{eval}\;\Varid{e}\mskip1.5mu]}
\]

\newcommand{\figcompiler}{
\begin{figure}
\abovedisplayskip=0pt
\belowdisplayskip=0pt
\raggedright
\begin{hscode}\SaveRestoreHook
\column{B}{@{}>{\hspre}l<{\hspost}@{}}%
\column{15}{@{}>{\hspre}l<{\hspost}@{}}%
\column{17}{@{}>{\hspre}l<{\hspost}@{}}%
\column{21}{@{}>{\hspre}l<{\hspost}@{}}%
\column{34}{@{}>{\hspre}l<{\hspost}@{}}%
\column{49}{@{}>{\hspre}l<{\hspost}@{}}%
\column{E}{@{}>{\hspre}l<{\hspost}@{}}%
\>[B]{}\textbf{\codefont module}\;\Conid{Compiler}\;\textbf{\codefont where}{}\<[E]%
\\[\blanklineskip]%
\>[B]{}\textbf{\codefont data}\;\Conid{Expr}\mathrel{=}\Conid{Val}\;\Conid{Int}\mid \Conid{Add}\;\Conid{Expr}\;\Conid{Expr}{}\<[E]%
\\[\blanklineskip]%
\>[B]{}\Varid{eval}\mathbin{::}\Conid{Expr}\mathbin{\text{\codefont ->}}\Conid{Int}{}\<[E]%
\\[-0.3ex]%
\>[B]{}\Varid{eval}\;(\Conid{Val}\;\Varid{n}){}\<[17]%
\>[17]{}\mathrel{=}\Varid{n}{}\<[E]%
\\[-0.3ex]%
\>[B]{}\Varid{eval}\;(\Conid{Add}\;\Varid{x}\;\Varid{y}){}\<[17]%
\>[17]{}\mathrel{=}\Varid{eval}\;\Varid{x}\mathbin{+}\Varid{eval}\;\Varid{y}{}\<[E]%
\\[\blanklineskip]%
\>[B]{}\textbf{\codefont type}\;\Conid{Stack}\mathrel{=}[\mskip1.5mu \Conid{Int}\mskip1.5mu]{}\<[E]%
\\[\blanklineskip]%
\>[B]{}\textbf{\codefont type}\;\Conid{Code}\mathrel{=}[\mskip1.5mu \Conid{Op}\mskip1.5mu]{}\<[E]%
\\[\blanklineskip]%
\>[B]{}\textbf{\codefont data}\;\Conid{Op}\mathrel{=}\Conid{PUSH}\;\Conid{Int}\mid \Conid{ADD}{}\<[E]%
\\[\blanklineskip]%
\>[B]{}\Varid{exec}\mathbin{::}\Conid{Code}\mathbin{\text{\codefont ->}}\Conid{Stack}\mathbin{\text{\codefont ->}}\Conid{Maybe}\;\Conid{Stack}{}\<[E]%
\\[-0.3ex]%
\>[B]{}\Varid{exec}\;[\mskip1.5mu \mskip1.5mu]\;{}\<[21]%
\>[21]{}\Varid{s}{}\<[34]%
\>[34]{}\mathrel{=}\Conid{Just}\;\Varid{s}{}\<[E]%
\\[-0.3ex]%
\>[B]{}\Varid{exec}\;(\Conid{PUSH}\;\Varid{n}{}\<[15]%
\>[15]{}\mathbin{:}\Varid{c})\;{}\<[21]%
\>[21]{}\Varid{s}{}\<[34]%
\>[34]{}\mathrel{=}\Varid{exec}\;\Varid{c}\;(\Varid{n}{}\<[49]%
\>[49]{}\mathbin{:}\Varid{s}){}\<[E]%
\\[-0.3ex]%
\>[B]{}\Varid{exec}\;(\Conid{ADD}{}\<[15]%
\>[15]{}\mathbin{:}\Varid{c})\;{}\<[21]%
\>[21]{}(\Varid{m}\mathbin{:}\Varid{n}\mathbin{:}\Varid{s}){}\<[34]%
\>[34]{}\mathrel{=}\Varid{exec}\;\Varid{c}\;(\Varid{n}\mathbin{+}\Varid{m}{}\<[49]%
\>[49]{}\mathbin{:}\Varid{s}){}\<[E]%
\\[-0.3ex]%
\>[B]{}\Varid{exec}\;(\Conid{ADD}{}\<[15]%
\>[15]{}\mathbin{:}\Varid{c})\;{}\<[21]%
\>[21]{}\anonymous {}\<[34]%
\>[34]{}\mathrel{=}\Conid{Nothing}{}\<[E]%
\\[\blanklineskip]%
\>[B]{}\Varid{comp}\mathbin{::}\Conid{Expr}\mathbin{\text{\codefont ->}}\Conid{Code}{}\<[E]%
\\[-0.3ex]%
\>[B]{}\Varid{comp}\;\Varid{e}\mathrel{=}\text{\codefont comp\textquotesingle}\;\Varid{e}\;[\mskip1.5mu \mskip1.5mu]{}\<[E]%
\\[\blanklineskip]%
\>[B]{}\text{\codefont comp\textquotesingle}\mathbin{::}\Conid{Expr}\mathbin{\text{\codefont ->}}\Conid{Code}\mathbin{\text{\codefont ->}}\Conid{Code}{}\<[E]%
\\[-0.3ex]%
\>[B]{}\text{\codefont comp\textquotesingle}\;(\Conid{Val}\;\Varid{n})\;{}\<[17]%
\>[17]{}\Varid{c}\mathrel{=}\Conid{PUSH}\;\Varid{n}\mathbin{:}\Varid{c}{}\<[E]%
\\[-0.3ex]%
\>[B]{}\text{\codefont comp\textquotesingle}\;(\Conid{Add}\;\Varid{x}\;\Varid{y})\;\Varid{c}\mathrel{=}\text{\codefont comp\textquotesingle}\;\Varid{x}\;(\text{\codefont comp\textquotesingle}\;\Varid{y}\;(\Conid{ADD}\mathbin{:}\Varid{c})){}\<[E]%
\ColumnHook
\end{hscode}\resethooks
\caption{Hutton's razor}
\label{fig:compiler}
\end{figure}}

\paragraph*{Experience}
Even in this simple example, the design of the compiler and its correctness
proof are subtle. In particular, in Hutton's original presentation, the \ensuremath{\Varid{exec}}
function is \emph{partial}: it does not handle stack underflow.  This partiality
guides Hutton's design; he presents and rejects an initial version of the
\ensuremath{\Varid{comp}} function because of this partiality.

Since Coq does not support partial functions, this posed an immediate problem.
This is why the code in \cref{fig:compiler} has been modified: we changed
 \ensuremath{\Varid{exec}} to return a \ensuremath{\Conid{Maybe}\;\Conid{Stack}}, not simply a \ensuremath{\Conid{Stack}}, and added the final
equation.  Once we made this small change and translated the code with
\texttt{hs-to-coq}, the proof of compiler correctness was easy.  In
fact, in Coq's interactive mode, users can follow the exact same (small) steps
of reasoning for this proof that Hutton provides in his textbook -- or use Coq’s proof automation to significantly speed up the proof process.

%

\figcompiler

\paragraph*{Conclusion}
We were successfully able to replicate a textbook correctness proof for a
Haskell programs, but along the way, we encountered the first significant
difference between Coq and Haskell, namely partiality (\cref{sec:partial}
provides more details).  Since we only set out to translate total code, we needed to
update the source code to be total; once we did so, we could translate the
textbook proofs to Coq directly.%
\asz{Note re. \cref{fig:compiler}: need to configure \texttt{lhs2\TeX} to output
  straight quotes, not curly quotes, for primes}

\subsection{Data Structure Correctness}\label{sec:bags}

\paragraph*{Objective}
In the last case study, we apply our tool to self-contained code that lives
within a large, existing code base. The \ensuremath{\Conid{Bag}} module%
\footnote{
    \url{http://git.haskell.org/ghc.git/blob/ghc-8.0.2-release:/compiler/utils/Bag.hs}}
from GHC~\cite{GHC} implements multisets with the following data type
declaration.
\begin{hscode}\SaveRestoreHook
\column{B}{@{}>{\hspre}l<{\hspost}@{}}%
\column{3}{@{}>{\hspre}c<{\hspost}@{}}%
\column{3E}{@{}l@{}}%
\column{6}{@{}>{\hspre}l<{\hspost}@{}}%
\column{8}{@{}>{\hspre}l<{\hspost}@{}}%
\column{E}{@{}>{\hspre}l<{\hspost}@{}}%
\>[B]{}\textbf{\codefont data}\;\Conid{Bag}\;\Varid{a}{}\<[E]%
\\[-0.3ex]%
\>[B]{}\hsindent{3}{}\<[3]%
\>[3]{}\mathrel{=}{}\<[3E]%
\>[6]{}\Conid{EmptyBag}{}\<[E]%
\\[-0.3ex]%
\>[B]{}\hsindent{3}{}\<[3]%
\>[3]{}\mid {}\<[3E]%
\>[6]{}\Conid{UnitBag}\;\Varid{a}{}\<[E]%
\\[-0.3ex]%
\>[B]{}\hsindent{3}{}\<[3]%
\>[3]{}\mid {}\<[3E]%
\>[6]{}\Conid{TwoBags}\;(\Conid{Bag}\;\Varid{a})\;(\Conid{Bag}\;\Varid{a}){}\<[E]%
\\[-0.3ex]%
\>[6]{}\hsindent{2}{}\<[8]%
\>[8]{}\mbox{\onelinecomment  INVARIANT: neither branch is empty}{}\<[E]%
\\[-0.3ex]%
\>[B]{}\hsindent{3}{}\<[3]%
\>[3]{}\mid {}\<[3E]%
\>[6]{}\Conid{ListBag}\;[\mskip1.5mu \Varid{a}\mskip1.5mu]{}\<[E]%
\\[-0.3ex]%
\>[6]{}\hsindent{2}{}\<[8]%
\>[8]{}\mbox{\onelinecomment  INVARIANT: the list is non-empty}{}\<[E]%
\ColumnHook
\end{hscode}\resethooks
The comments in this declaration specify the two invariants that a value of
this type must satisfy.  Furthermore, at the top of the file, the
documentation gives the intended semantics of this type: a \ensuremath{\Conid{Bag}} is ``an
unordered collection with duplicates''.  In fact, the current implementation
satisfies the stronger property that all operations on \ensuremath{\Conid{Bag}}s preserve the
\emph{order} of elements, so we can say that their semantics is given by the
function \ensuremath{\Varid{bagToList}\mathbin{::}\Conid{Bag}\;\Varid{a}\mathbin{\text{\codefont ->}}[\mskip1.5mu \Varid{a}\mskip1.5mu]}, which is defined in the module.

\paragraph*{Experience}

The part of the module that we are interested in is fairly straightforward; in
addition to the \ensuremath{\Conid{Bag}} type, it contains a number of basic functions, such as
\begin{hscode}\SaveRestoreHook
\column{B}{@{}>{\hspre}l<{\hspost}@{}}%
\column{13}{@{}>{\hspre}l<{\hspost}@{}}%
\column{E}{@{}>{\hspre}l<{\hspost}@{}}%
\>[B]{}\Varid{isEmptyBag}{}\<[13]%
\>[13]{}\mathbin{::}\Conid{Bag}\;\Varid{a}\mathbin{\text{\codefont ->}}\Conid{Bool}{}\<[E]%
\\[-0.3ex]%
\>[B]{}\Varid{unionBags}{}\<[13]%
\>[13]{}\mathbin{::}\Conid{Bag}\;\Varid{a}\mathbin{\text{\codefont ->}}\Conid{Bag}\;\Varid{a}\mathbin{\text{\codefont ->}}\Conid{Bag}\;\Varid{a}{}\<[E]%
\ColumnHook
\end{hscode}\resethooks

We formalize the combined invariants as a boolean predicate
\ensuremath{\Varid{well\char95 formed\char95 bag}}\iflong\asz{Show code?}\fi.  Then, for each translated
function,\footnote{We skipped monadic functions such as \ensuremath{\Varid{mapBagM}}, along with
  three further functions that referred to code we did not translate.} we prove
up to two theorems:
\begin{enumerate}
\item We prove that each function is equivalent, with respect to \ensuremath{\Varid{bagToList}}, to
  the corresponding list function.
\item If the function returns a \ensuremath{\Conid{Bag}}, we prove that it preserves the \ensuremath{\Conid{Bag}}
  invariants.
\end{enumerate}
Thus, for example, we prove the following three theorems about \ensuremath{\Varid{isEmptyBag}} and
\ensuremath{\Varid{unionBags}}:
\begin{hscode}\SaveRestoreHook
\column{B}{@{}>{\hspre}l<{\hspost}@{}}%
\column{3}{@{}>{\hspre}l<{\hspost}@{}}%
\column{E}{@{}>{\hspre}l<{\hspost}@{}}%
\>[B]{}{\textbf{\codefont Theorem}}\;\Varid{isEmptyBag\char95 ok}\;\mathord{\text{\codefont \{}}\Conid{A}\mathord{\text{\codefont \}}}\;(\Varid{b}\mathbin{:}\Conid{Bag}\;\Conid{A})\mathbin{:}{}\<[E]%
\\[-0.3ex]%
\>[B]{}\hsindent{3}{}\<[3]%
\>[3]{}\Varid{well\char95 formed\char95 bag}\;\Varid{b}\mathbin{\text{\codefont ->}}{}\<[E]%
\\[-0.3ex]%
\>[B]{}\hsindent{3}{}\<[3]%
\>[3]{}\Varid{isEmptyBag}\;\Varid{b}\mathrel{=}\Varid{null}\;(\Varid{bagToList}\;\Varid{b})\mathop{\text{.}}{}\<[E]%
\\[\blanklineskip]%
\>[B]{}{\textbf{\codefont Theorem}}\;\Varid{unionBags\char95 ok}\;\mathord{\text{\codefont \{}}\Conid{A}\mathord{\text{\codefont \}}}\;(\Varid{b1}\;\Varid{b2}\mathbin{:}\Conid{Bag}\;\Conid{A})\mathbin{:}{}\<[E]%
\\[-0.3ex]%
\>[B]{}\hsindent{3}{}\<[3]%
\>[3]{}\Varid{bagToList}\;(\Varid{unionBags}\;\Varid{b1}\;\Varid{b2})\mathrel{=}{}\<[E]%
\\[-0.3ex]%
\>[B]{}\hsindent{3}{}\<[3]%
\>[3]{}\Varid{bagToList}\;\Varid{b1}\mathbin{\text{\codefont ++}}\Varid{bagToList}\;\Varid{b2}\mathop{\text{.}}{}\<[E]%
\\[\blanklineskip]%
\>[B]{}{\textbf{\codefont Theorem}}\;\Varid{unionBags\char95 wf}\;\mathord{\text{\codefont \{}}\Conid{A}\mathord{\text{\codefont \}}}\;(\Varid{b1}\;\Varid{b2}\mathbin{:}\Conid{Bag}\;\Conid{A})\mathbin{:}{}\<[E]%
\\[-0.3ex]%
\>[B]{}\hsindent{3}{}\<[3]%
\>[3]{}\Varid{well\char95 formed\char95 bag}\;\Varid{b1}\mathbin{\text{\codefont ->}}\Varid{well\char95 formed\char95 bag}\;\Varid{b2}\mathbin{\text{\codefont ->}}{}\<[E]%
\\[-0.3ex]%
\>[B]{}\hsindent{3}{}\<[3]%
\>[3]{}\Varid{well\char95 formed\char95 bag}\;(\Varid{unionBags}\;\Varid{b1}\;\Varid{b2})\mathop{\text{.}}{}\<[E]%
\ColumnHook
\end{hscode}\resethooks
Interestingly, we can see that \ensuremath{\Varid{isEmptyBag}}'s correctness theorem requires that
its argument satisfy the \ensuremath{\Conid{Bag}} invariants, but \ensuremath{\Varid{unionBags}}'s does not.

\sloppy

\subparagraph*{Verifying \ensuremath{\Conid{Bag}}}

The verification effort proceeded just as though we were verifying any data
structure library written in Coq.  We verified nineteen different functions on
\ensuremath{\Conid{Bag}}s, and no proof was longer than eight lines (using the \texttt{ssreflect}
tactic library~\cite{ssreflect}).

\fussy

Along the way, we discovered a minor omission in the documentation of the
\ensuremath{\Varid{foldBag}} function. This function has type
\begin{hscode}\SaveRestoreHook
\column{B}{@{}>{\hspre}l<{\hspost}@{}}%
\column{E}{@{}>{\hspre}l<{\hspost}@{}}%
\>[B]{}\Varid{foldBag}\mathbin{::}(\Varid{r}\mathbin{\text{\codefont ->}}\Varid{r}\mathbin{\text{\codefont ->}}\Varid{r})\mathbin{\text{\codefont ->}}(\Varid{a}\mathbin{\text{\codefont ->}}\Varid{r})\mathbin{\text{\codefont ->}}\Varid{r}\mathbin{\text{\codefont ->}}\Conid{Bag}\;\Varid{a}\mathbin{\text{\codefont ->}}\Varid{r}{}\<[E]%
\ColumnHook
\end{hscode}\resethooks
The expression \ensuremath{\Varid{foldBag}\;\Varid{t}\;\Varid{u}\;\Varid{e}} maps \ensuremath{\Varid{u}} over every element of the bag and then,
starting with \ensuremath{\Varid{e}}, combines these results from the right using the operator \ensuremath{\Varid{t}},
à la \ensuremath{\Varid{foldr}}.

The documentation for \ensuremath{\Varid{foldBag}} requires that \ensuremath{\Varid{t}} be associative, and says that
it is then a ``more tail-recursive'' version of a commented-out reference
implementation which combines the results according the internal structure of
the \ensuremath{\Conid{Bag}} instead of from the right.  However, as we discovered when attempting
to prove the two implementations equal, the reference implementation is
\emph{not} the same as \ensuremath{\Varid{foldBag}} in all cases -- they are only the same when \ensuremath{\Varid{e}}
is the identity for \ensuremath{\Varid{t}}.  This discrepancy is minor, but has been present for
over~21~years~\cite{ghc-foldBag-commit}.

\subparagraph*{Selectively translating  \ensuremath{\Conid{Bag}}}

As a part of GHC, the \ensuremath{\Conid{Bag}} module cannot stand on its own; it imports a number
of other modules from GHC, such as \ensuremath{\Conid{Outputable}} and \ensuremath{\Conid{Util}}.  However, there is a
great deal of code we don't care about in GHC.  For example, the
\ensuremath{\Conid{Outputable}} module contains infrastructure for pretty printing. For our
verification goals, this module is completely irrelevant, so it would be
unfortunate if we could not proceed without translating it into Coq. But it
would be equally unfortunate if we had to edit the GHC sources to remove code
that we were not interested in.

It is for these sorts of reasons that \texttt{hs-to-coq} supports declaratively
configuring the translation process: it can take as input a file of declarative
instructions, called \emph{edits}, that influence the translation process.  One
such instruction is to skip translating a module:
\begin{myhscode}%
\defaultcolumn{@{}>{\codefont}l@{\ }}%
\> {\codefont\textbf{skip module}} Outputable
\end{myhscode}
Similar instructions exist to skip functions, type classes, instances and type
class methods; for example, the \ensuremath{\Conid{Util}} module contains a number of utility
functions that aren't used by \ensuremath{\Conid{Bag}}, and so are unnecessary.

\paragraph*{Conclusion}
Because \texttt{hs-to-coq}’s translation is configurable, we were able to
slice the code of interest out of a large, existing codebase, without having to
translate irrelevant parts or change the original source code. Once translated,
the code was pleasant and straightforward to work with, and we completed both
invariant preservation and semantic correctness proofs.  We also saw that
specifications are subtle, and edge cases in \emph{documentation} can be caught
by such verification.

\section{The Design and Implementation of \texttt{hs-to-coq}}
\label{sec:design}
\label{sec:challenges}

The previous section describes \texttt{hs-to-coq} in action: it processes a
Haskell program, along with a separate files of ``\mbox{edits}'', which are commands
that modify the translation in well-defined ways, and produces verifiable Coq
code. Our design goals for \texttt{hs-to-coq} include:
\begin{enumerate}
\item Produce output resembling the original input;
\item Produce output amenable to interactive proof development;
\item Handle features commonly found in modern Haskell developments; and
\item Apply to source code as is, even if it is part of a larger development.
\end{enumerate}
We have made the somewhat controversial choice to focus on \emph{total} Haskell
programs. This choice follows from our first two goals above: total programs
require fewer modifications to be accepted by Coq (for example, there is no need
to use a monad to model partiality) and provide more assurances (if a
translation is successful we know that the code is total). At the same time,
reasoning about total functions is simpler than reasoning about partial ones, so
we encourage Haskell proof development by concentrating on this domain.

The configurable edits support this design.  Example edits include
skipping functions that aren't being verified, or renaming a translated type
or value to its Coq equivalent for interoperability.  By providing this in a
separate file, this per-project changes do not need to be applied to the code
itself, and do not have to be re-done as the code evolves.

We use the Glasgow Haskell Compiler (GHC), version 8.0.2, as a library~\cite{GHC}.
By using its parser, \texttt{hs-to-coq} can process most Haskell code as seen in the wild.
In fact, our tool adopts the first two stages of GHC\@. First, the source code
passes through the \emph{parser} and an AST is produced.  This AST then goes
through the \emph{renamer}, which resolves name references and ensures that
programs are well scoped. Based on this, the tool generates the Coq output.

Note that \texttt{hs-to-coq} generates the Coq output before the
\emph{typechecking} and \emph{desugaring} phases. Going after the desugaring, and hence translating GHC’s intermediate language Core, would certainly simplify the translation. But the resulting code would look too different from the Haskell source code, and go against our first goal.

Many of the syntactic constructs found in Haskell have direct equivalents in
Coq: algebraic data types, function definitions, basic pattern matching,
function application, let-bindings, and so on.  Translating these constructs is
immediate.  Other syntactic constructs may not exist in Coq, but are
straightforward to desugar: \ensuremath{\textbf{\codefont where}} clauses become \ensuremath{{\textbf{\codefont match}}} or
\ensuremath{\textbf{\codefont let}} expressions, \ensuremath{\textbf{\codefont do}} notation and list comprehensions turn into explicit
function calls, etc.

However, many complex Haskell features do not map so cleanly
onto Coq features. In the following we discuss our resolution of these challenging translations in the context of our design goals.

\subsection{Module System}
\label{sec:modules}

Haskell and Coq have wildly different approaches to their module systems, but
thankfully they both have one.  The largest point of commonality is that in both
Haskell and Coq, each source file creates a single module, with its name
determined by the file name and the path thereto.  The method for handling
modules is thus twofold:
\begin{itemize}
\item translate each Haskell file into a distinct Coq file; and
\item always refer to all names fully qualified to avoid any differences between
  the module systems.
\end{itemize}

In each Coq module, we make available (through \ensuremath{{\textbf{\codefont Require}}}) all modules that are
referenced by any identifiers.  We do this instead of translating the Haskell
\ensuremath{\textbf{\codefont import}} statements directly because of one of the differences between Haskell
and Coq: Haskell allows a module to \emph{re-export} identifiers that it
imported, but GHC's frontend only keeps track of the \emph{original} module's
name.  So the fully-qualified name we generate refers to something further
back in the module tree that must itself be imported.%
\scw{Should we say why we don't \ensuremath{{\textbf{\codefont Import}}} everything and produce fully
  qualified names. Or why we generate \ensuremath{\Varid{op\char95 zeze\char95 \char95 }} instead of using the
  notation?}

\subsection{Records}\label{sec:records}
In Haskell, data types can be defined as \emph{records}. For example, the definition of the functions \ensuremath{\Varid{getCurrent}} and \ensuremath{\Varid{getSuccs}} in \cref{fig:succs} could be omitted if the data type were defined as
\begin{hscode}\SaveRestoreHook
\column{B}{@{}>{\hspre}l<{\hspost}@{}}%
\column{E}{@{}>{\hspre}l<{\hspost}@{}}%
\>[B]{}\textbf{\codefont data}\;\Conid{Succs}\;\Varid{a}\mathrel{=}\Conid{Succs}\;\mathord{\text{\codefont \{}}\Varid{getCurrent}\mathbin{::}\Varid{a},\Varid{getSuccs}\mathbin{::}[\mskip1.5mu \Varid{a}\mskip1.5mu]\mathord{\text{\codefont \}}}{}\<[E]%
\ColumnHook
\end{hscode}\resethooks
The type is the same, but naming the fields enables some extra features:
(1)~unordered value creation,
(2)~named pattern matching,
(3)~field accessors, and
(4)~field updates~\cite{haskell2010}.
In addition, with GHC extensions, it also enables
(5) \emph{record wild cards}: a pattern or expression of the form
\ensuremath{\Conid{Succs}\;\mathord{\text{\codefont \{}}\mathinner{\ldotp\ldotp}\mathord{\text{\codefont \}}}} binds each field to a variable of the same name.

Coq features support for single-constructor records that can do (1--3), although
with minor differences; however, it lacks support for (4--5).  More importantly,
however, Haskell records are \emph{per-constructor} -- a sum type can contain
fields for each of its constructors.  Coq does not support this at all.
Consequently, \texttt{hs-to-coq} keeps track of record field names
during the translation process.  Constructors with record fields are translated as though
they had no field names, and the Coq accessor functions are generated
separately.  During pattern matching or updates -- particularly with wild cards
-- the field names are linked to the appropriate positional field.


\newcommand{\metavar}[1]{\textit{#1}}
\newcommand{\metafun}[1]{\textit{\textsf{#1}}}

\newcommand{\figguardsyntax}{%
\begin{figure}
\abovedisplayskip=0pt
\belowdisplayskip=0pt
\[
\begin{array}{r@{\,}c@{\,}ll}
\metavar e
& \Coloneqq & \metavar x                                  & \text{variables} \\
& \mid      & \ensuremath{\textbf{\codefont case}}~\metavar e~\ensuremath{\textbf{\codefont of}}~\metavar{alt}^+      & \text{case analysis} \\
& \mid      & \ensuremath{\textbf{\codefont let}}~\metavar{bind}^+~\ensuremath{\textbf{\codefont in}}~\metavar e      & \text{let binding} \\
& \mid      & \bot                                        & \text{partiality marker}\\
[1ex]
\metavar{bind}
& \Coloneqq & \metavar x = \metavar e                     & \text{binding} \\
[1ex]
\metavar{alt}
& \Coloneqq & \metavar{pat}~\metavar{grhs}^+~[\ensuremath{\textbf{\codefont where}}~\metavar{bind}^+]
                                                          & \text{alternative} \\
[1ex]
\metavar{pat}
& \Coloneqq & \metavar{con}~\metavar{pat}^\ast            & \text{constructor pattern} \\
& \mid      & \metavar x                                  & \text{variable pattern} \\
& \mid      & \_                                          & \text{wild pattern} \\
[1ex]
\metavar{grhs}
& \Coloneqq & \ensuremath{\mathbin{\text{\codefont ->}}} \metavar e                             & \text{unguarded rhs} \\
& \mid      & \metavar{guard}~\metavar grhs               & \text{guarded rhs} \\
[1ex]
\metavar{guard}
& \Coloneqq & \metavar e                                  & \text{boolean guard} \\
& \mid      & \metavar{pat} \ensuremath{\leftarrow } \metavar e               & \text{pattern guard} \\
& \mid      & \ensuremath{\textbf{\codefont let}}~\metavar{binds}^+                     & \text{binding guard}
\end{array}
\]

\caption{An abstract grammar for pattern matches with guards}
\label{fig:guard-syntax}
\end{figure}
}

\newcommand{\figguardalgo}{%
\begin{figure*}
\abovedisplayskip=0pt
\belowdisplayskip=0pt
\begin{align*}
\ensuremath{\textbf{\codefont case}}~\metavar e~\ensuremath{\textbf{\codefont of}}~\metavar{alt}^+
&\rightsquigarrow
\ensuremath{\textbf{\codefont let}}~\ensuremath{\Varid{scrut}}=\metavar e~\ensuremath{\textbf{\codefont in}}
\\&\phantom{\rightsquigarrow{}}
\metafun{splitAlts}(\metavar{alt}^+, \bot)
\\[1ex]
\metafun{splitAlts}(,\metavar{def})
&\rightsquigarrow \metavar{def} \\
\metafun{splitAlts}(\metavar{a}_1\ldots\metavar{a}_n, \metavar{def})
&\rightsquigarrow
\ensuremath{\textbf{\codefont let}}~\ensuremath{\Varid{k}}=
\metafun{mkCase}(\metavar{a}_j\ldots\metavar{a}_n, \metavar{def})~\ensuremath{\textbf{\codefont in}}
\\&\phantom{\rightsquigarrow{}}
\metafun{splitAlts}(\metavar{a}_1\ldots\metavar{a}_{j-1}, \ensuremath{\Varid{k}})
\\&\phantom{\rightsquigarrow{}}
\text{\quad for a (small) $j$ for which $\metavar a_j,\ldots,\metavar a_n$ are mutually exclusive}
\\[1ex]
\metafun{mkCase}(\metavar{a}_1\ldots\metavar{a}_n, \metavar{def})
&\rightsquigarrow%
\ensuremath{\textbf{\codefont case}}~\ensuremath{\Varid{scrut}}~\ensuremath{\textbf{\codefont of}}~\begin{array}[t]{@{}l@{\,}c@{\,}l}%
                    \metafun{altPat}(\metavar a_1) &\ensuremath{\mathbin{\text{\codefont ->}}}&
                    \metafun{altRHS}(\metavar a_1, \metavar{def})\\
                    & \vdots &\\
                    \metafun{altPat}(\metavar a_n) &\ensuremath{\mathbin{\text{\codefont ->}}}&
                    \metafun{altRHS}(\metavar a_n, \metavar{def})\\
                    \end{array}
\\&\phantom{\rightsquigarrow{}}
\text{\quad if $\metafun{altPat}(\metavar a_1),\ldots,\metafun{altPat}(\metavar a_n)$ are complete}\\
\metafun{mkCase}(\metavar{a}_1\ldots\metavar{a}_n, \metavar{def})
&\rightsquigarrow%
\ensuremath{\textbf{\codefont case}}~\ensuremath{\Varid{scrut}}~\ensuremath{\textbf{\codefont of}}~\begin{array}[t]{@{}l@{\,}c@{\,}l}%
                    \metafun{altPat}(\metavar a_1) & \ensuremath{\mathbin{\text{\codefont ->}}} &
                    \metafun{altRHS}(\metavar a_1, \metavar{def})\\
                    &\vdots&\\
                    \metafun{altPat}(\metavar a_n) & \ensuremath{\mathbin{\text{\codefont ->}}} &
                    \metafun{altRHS}(\metavar a_n, \metavar{def})\\
                    \_ &\ensuremath{\mathbin{\text{\codefont ->}}}& \metavar{def}
                    \end{array}
\\&\phantom{\rightsquigarrow{}}
\text{\quad otherwise}
\\[1ex]
\metafun{altPat}(\metavar{pat}~\metavar{grhs}^+~\ensuremath{\textbf{\codefont where}}~\metavar{bind}^+)
&\rightsquigarrow%
\metavar{pat}
\\[1ex]
\metafun{altRHS}(\metavar{pat}~\metavar{grhs}^+~\ensuremath{\textbf{\codefont where}}~\metavar{bind}^+, \metavar{def})
&\rightsquigarrow%
\begin{array}[t]{@{}l}
\ensuremath{\textbf{\codefont let}}~\metavar{bind}^+~\ensuremath{\textbf{\codefont in}}\\
\metafun{mkG}(\metavar{grhs}^+, \metavar{def})
\end{array}
\\
\metafun{altRHS}(\metavar{pat}~\metavar{grhs}^+, \metavar{def})
&\rightsquigarrow%
\metafun{mkGs}(\metavar{grhs}^+, \metavar{def})
\\[1ex]
\metafun{mkGs}(\metavar{grhs}, \metavar{def})
&\rightsquigarrow%
\metafun{mkG}(\metavar{grhs}, \metavar{def})
\\
\metafun{mkGs}(\metavar{grhs}^+~\metavar{grhs}, \metavar{def})
&\rightsquigarrow%
\begin{array}[t]{@{}l}
\ensuremath{\textbf{\codefont let}}~\ensuremath{\Varid{k}}=\metafun{mkG}(\metavar{grhs}, \metavar{def})~\ensuremath{\textbf{\codefont in}} \\
\metafun{mkGs}(\metavar{grhs}^+, \ensuremath{\Varid{k}})
\end{array}
\\[1ex]
\metafun{mkG}(\ensuremath{\mathbin{\text{\codefont ->}}} \metavar{e}, \metavar{def})
&\rightsquigarrow%
\metavar{e} \\
\metafun{mkG}(\metavar{e}~\metavar{grhs}, \metavar{def})
&\rightsquigarrow%
\ensuremath{\textbf{\codefont if}}~e~\ensuremath{\textbf{\codefont then}}~\metafun{mkG}(\metavar{grhs}, \metavar{def})~\ensuremath{\textbf{\codefont else}}~\metavar{def}\\
\metafun{mkG}((\metavar{pat} \ensuremath{\leftarrow }\metavar{e})~\metavar{grhs}, \metavar{def})
&\rightsquigarrow%
\ensuremath{\textbf{\codefont case}}~e~\ensuremath{\textbf{\codefont of}}~\begin{array}[t]{@{}l@{\,}l}
              \metavar{pat}~&\ensuremath{\mathbin{\text{\codefont ->}}} \metafun{mkG}(\metavar{grhs}, \metavar{def}) \\
              \_            &\ensuremath{\mathbin{\text{\codefont ->}}} \metavar{def}
              \end{array} \\
\metafun{mkG}((\ensuremath{\textbf{\codefont let}}~\metavar{binds}^+)~\metavar{grhs}, \metavar{def})
&\rightsquigarrow%
\ensuremath{\textbf{\codefont let}}~\metavar{binds}^+~\ensuremath{\textbf{\codefont in}}~\metafun{mkG}(\metavar{grhs}, \metavar{def})
\end{align*}
\caption{Translation of case expressions with guards}
\label{fig:guard-algo}
\end{figure*}
}

\subsection{Patterns in Function Definitions}

Haskell function definitions allow the programmer to have patterns as
parameters:
\begin{hscode}\SaveRestoreHook
\column{B}{@{}>{\hspre}l<{\hspost}@{}}%
\column{E}{@{}>{\hspre}l<{\hspost}@{}}%
\>[B]{}\Varid{uncurry}\mathbin{::}(\Varid{a}\mathbin{\text{\codefont ->}}\Varid{b}\mathbin{\text{\codefont ->}}\Varid{c})\mathbin{\text{\codefont ->}}(\Varid{a},\Varid{b})\mathbin{\text{\codefont ->}}\Varid{c}{}\<[E]%
\\[-0.3ex]%
\>[B]{}\Varid{uncurry}\;\Varid{f}\;(\Varid{x},\Varid{y})\mathrel{=}\Varid{f}\;\Varid{x}\;\Varid{y}{}\<[E]%
\ColumnHook
\end{hscode}\resethooks
This code is not allowed in Coq; pattern matching is only performed by the \ensuremath{{\textbf{\codefont match}}}
expression.  Instead, programmers first have to name the parameter, and then
perform a separate pattern match:
\begin{hscode}\SaveRestoreHook
\column{B}{@{}>{\hspre}l<{\hspost}@{}}%
\column{3}{@{}>{\hspre}l<{\hspost}@{}}%
\column{5}{@{}>{\hspre}l<{\hspost}@{}}%
\column{7}{@{}>{\hspre}l<{\hspost}@{}}%
\column{E}{@{}>{\hspre}l<{\hspost}@{}}%
\>[B]{}{\textbf{\codefont Definition}}\;\Varid{uncurry}\;\mathord{\text{\codefont \{}}\Varid{a}\mathord{\text{\codefont \}}}\;\mathord{\text{\codefont \{}}\Varid{b}\mathord{\text{\codefont \}}}\;\mathord{\text{\codefont \{}}\Varid{c}\mathord{\text{\codefont \}}}\mathbin{:}{}\<[E]%
\\[-0.3ex]%
\>[B]{}\hsindent{3}{}\<[3]%
\>[3]{}(\Varid{a}\mathbin{\text{\codefont ->}}\Varid{b}\mathbin{\text{\codefont ->}}\Varid{c})\mathbin{\text{\codefont ->}}\Varid{a}\mathbin{*}\Varid{b}\mathbin{\text{\codefont ->}}\Varid{c}\mathbin{:=}{}\<[E]%
\\[-0.3ex]%
\>[B]{}\hsindent{3}{}\<[3]%
\>[3]{}{\textbf{\codefont fun}}\;\Varid{arg\char95 10\char95 \char95 }\;\Varid{arg\char95 11\char95 \char95 }\mathbin{\text{\codefont =>}}{}\<[E]%
\\[-0.3ex]%
\>[3]{}\hsindent{2}{}\<[5]%
\>[5]{}{\textbf{\codefont match}}\;\Varid{arg\char95 10\char95 \char95 },\Varid{arg\char95 11\char95 \char95 }\;{\textbf{\codefont with}}{}\<[E]%
\\[-0.3ex]%
\>[5]{}\hsindent{2}{}\<[7]%
\>[7]{}\mid \Varid{f},\Varid{pair}\;\Varid{x}\;\Varid{y}\mathbin{\text{\codefont =>}}\Varid{f}\;\Varid{x}\;\Varid{y}{}\<[E]%
\\[-0.3ex]%
\>[3]{}\hsindent{2}{}\<[5]%
\>[5]{}{\textbf{\codefont end}}\mathop{\text{.}}{}\<[E]%
\ColumnHook
\end{hscode}\resethooks

This translation extends naturally to functions that are defined using
multiple equations, as seen in the \ensuremath{\Varid{map}} function in \cref{sec:introduction}.

\subsection{Pattern Matching With Guards}
\label{sec:guards}

Another pattern-related challenge is posed by \emph{guards}, and translation
tools similar to ours have gotten their semantics wrong (see \cref{sec:related}).

Guards are side conditions that can be attached to a function equation or a
\ensuremath{\textbf{\codefont case}} alternative.  If the pattern matches, but the condition is not satisfied,
then the next equation is tried. A typical example is the \ensuremath{\Varid{take}}
function from the Haskell standard library, where \ensuremath{\Varid{take}\;\Varid{n}\;\Varid{xs}} returns the first
\ensuremath{\Varid{n}} elements of \ensuremath{\Varid{xs}}:
\begin{hscode}\SaveRestoreHook
\column{B}{@{}>{\hspre}l<{\hspost}@{}}%
\column{7}{@{}>{\hspre}l<{\hspost}@{}}%
\column{10}{@{}>{\hspre}l<{\hspost}@{}}%
\column{17}{@{}>{\hspre}l<{\hspost}@{}}%
\column{27}{@{}>{\hspre}c<{\hspost}@{}}%
\column{27E}{@{}l@{}}%
\column{30}{@{}>{\hspre}l<{\hspost}@{}}%
\column{E}{@{}>{\hspre}l<{\hspost}@{}}%
\>[B]{}\Varid{take}\mathbin{::}\Conid{Int}\mathbin{\text{\codefont ->}}[\mskip1.5mu \Varid{a}\mskip1.5mu]\mathbin{\text{\codefont ->}}[\mskip1.5mu \Varid{a}\mskip1.5mu]{}\<[E]%
\\[-0.3ex]%
\>[B]{}\Varid{take}\;{}\<[7]%
\>[7]{}\Varid{n}\;{}\<[10]%
\>[10]{}\anonymous {}\<[17]%
\>[17]{}\mid \Varid{n}\mathbin{\text{\codefont <=}}\mathrm{0}{}\<[27]%
\>[27]{}\mathrel{=}{}\<[27E]%
\>[30]{}[\mskip1.5mu \mskip1.5mu]{}\<[E]%
\\[-0.3ex]%
\>[B]{}\Varid{take}\;{}\<[7]%
\>[7]{}\anonymous \;{}\<[10]%
\>[10]{}[\mskip1.5mu \mskip1.5mu]{}\<[27]%
\>[27]{}\mathrel{=}{}\<[27E]%
\>[30]{}[\mskip1.5mu \mskip1.5mu]{}\<[E]%
\\[-0.3ex]%
\>[B]{}\Varid{take}\;{}\<[7]%
\>[7]{}\Varid{n}\;{}\<[10]%
\>[10]{}(\Varid{x}\mathbin{:}\Varid{xs}){}\<[27]%
\>[27]{}\mathrel{=}{}\<[27E]%
\>[30]{}\Varid{x}\mathbin{:}\Varid{take}\;(\Varid{n}\mathbin{-}\mathrm{1})\;\Varid{xs}{}\<[E]%
\ColumnHook
\end{hscode}\resethooks
The patterns in the first equation match any argument; however, the match only
succeeds if \ensuremath{\Varid{n}\mathbin{\text{\codefont <=}}\mathrm{0}} as well.  If \ensuremath{\Varid{n}} is positive, that equation is skipped, and
pattern matching proceeds to the next two equations.

Guards occur in three variants:
\begin{enumerate}
\item A \emph{boolean guard} is an expression \ensuremath{\Varid{expr}} of type \ensuremath{\Conid{Bool}}, as we saw
  in \ensuremath{\Varid{take}}. It succeeds if \ensuremath{\Varid{expr}} evaluates to \ensuremath{\Conid{True}}.
\item A \emph{pattern guard} is of the form \ensuremath{\Varid{pat}\leftarrow \Varid{expr}}.  It succeeds if the
  expression \ensuremath{\Varid{expr}} matches the pattern \ensuremath{\Varid{pat}}, and brings the variables in \ensuremath{\Varid{pat}}
  into scope just as any pattern match would.
\item A \emph{local declaration} of the form \ensuremath{\textbf{\codefont let}\;\Varid{x}\mathrel{=}\Varid{e}}.  This binds \ensuremath{\Varid{x}} to
  \ensuremath{\Varid{e}}, bringing \ensuremath{\Varid{x}} into scope, and always succeeds.
\end{enumerate}
Each equation can be guarded by a multiple guards, separated by commas, all of which must succeed in turn for this equation to be used.

Coq does not support guards, so \texttt{hs-to-coq}'s translation has to
eliminate them.  Conveniently, the Haskell Report~\cite{haskell2010} defines the
semantics of pattern matching with guards in terms of a sequence of rewrites, at
the end of which all guards have been removed and all case expressions are of
the following, primitive form:
\begin{hscode}\SaveRestoreHook
\column{B}{@{}>{\hspre}l<{\hspost}@{}}%
\column{12}{@{}>{\hspre}l<{\hspost}@{}}%
\column{25}{@{}>{\hspre}l<{\hspost}@{}}%
\column{E}{@{}>{\hspre}l<{\hspost}@{}}%
\>[B]{}\textbf{\codefont case}\;\Varid{e}\;\textbf{\codefont of}\;{}\<[12]%
\>[12]{}\Conid{K}\;\Varid{x1}\;{\ldots}\;\Varid{xN}{}\<[25]%
\>[25]{}\mathbin{\text{\codefont ->}}\Varid{e1}{}\<[E]%
\\[-0.3ex]%
\>[12]{}\anonymous {}\<[25]%
\>[25]{}\mathbin{\text{\codefont ->}}\Varid{e2}{}\<[E]%
\ColumnHook
\end{hscode}\resethooks
According to these rules, the \ensuremath{\Varid{take}} function defined above would be translated
to something like
\begin{hscode}\SaveRestoreHook
\column{B}{@{}>{\hspre}l<{\hspost}@{}}%
\column{14}{@{}>{\hspre}l<{\hspost}@{}}%
\column{20}{@{}>{\hspre}l<{\hspost}@{}}%
\column{24}{@{}>{\hspre}c<{\hspost}@{}}%
\column{24E}{@{}l@{}}%
\column{28}{@{}>{\hspre}l<{\hspost}@{}}%
\column{34}{@{}>{\hspre}c<{\hspost}@{}}%
\column{34E}{@{}l@{}}%
\column{38}{@{}>{\hspre}l<{\hspost}@{}}%
\column{E}{@{}>{\hspre}l<{\hspost}@{}}%
\>[B]{}\Varid{take}\mathbin{::}\Conid{Int}\mathbin{\text{\codefont ->}}[\mskip1.5mu \Varid{a}\mskip1.5mu]\mathbin{\text{\codefont ->}}[\mskip1.5mu \Varid{a}\mskip1.5mu]{}\<[E]%
\\[-0.3ex]%
\>[B]{}\Varid{take}\;\Varid{n}\;\Varid{xs}\mathrel{=}{}\<[14]%
\>[14]{}\textbf{\codefont if}\;\Varid{n}\mathbin{\text{\codefont <=}}\mathrm{0}{}\<[E]%
\\[-0.3ex]%
\>[14]{}\textbf{\codefont then}\;[\mskip1.5mu \mskip1.5mu]{}\<[E]%
\\[-0.3ex]%
\>[14]{}\textbf{\codefont else}\;{}\<[20]%
\>[20]{}\textbf{\codefont case}\;\Varid{xs}\;\textbf{\codefont of}{}\<[E]%
\\[-0.3ex]%
\>[20]{}[\mskip1.5mu \mskip1.5mu]{}\<[24]%
\>[24]{}\mathbin{\text{\codefont ->}}{}\<[24E]%
\>[28]{}[\mskip1.5mu \mskip1.5mu]{}\<[E]%
\\[-0.3ex]%
\>[20]{}\anonymous {}\<[24]%
\>[24]{}\mathbin{\text{\codefont ->}}{}\<[24E]%
\>[28]{}\textbf{\codefont case}\;\Varid{xs}\;\textbf{\codefont of}{}\<[E]%
\\[-0.3ex]%
\>[28]{}\Varid{x}\mathbin{:}\Varid{xs}{}\<[34]%
\>[34]{}\mathbin{\text{\codefont ->}}{}\<[34E]%
\>[38]{}\Varid{x}\mathbin{:}\Varid{take}\;(\Varid{n}\mathbin{-}\mathrm{1})\;\Varid{xs}{}\<[E]%
\\[-0.3ex]%
\>[28]{}\anonymous {}\<[34]%
\>[34]{}\mathbin{\text{\codefont ->}}{}\<[34E]%
\>[38]{}\Varid{error}\;\text{\itshape\texttt{\char34}No~match\texttt{\char34}}{}\<[E]%
\ColumnHook
\end{hscode}\resethooks

Unfortunately, this approach is unsuitable for \texttt{hs-to-coq} as the final
pattern match in this sequence requires an catch-all case to be complete.  This
requires an expression of arbitrary type, which exists in Haskell (\ensuremath{\Varid{error}\;{\ldots}\;}),
but cannot exist in Coq.  Additionally, since Coq supports nested patterns (such
as \ensuremath{\Conid{Just}\;(\Varid{x}\mathbin{:}\Varid{xs})}), we want to preserve them when translating Haskell code.

Therefore, we are more careful when translating case expressions with guards,
and we keep mutually exclusive patterns within the same \ensuremath{{\textbf{\codefont match}}}. This way, the
translated \ensuremath{\Varid{take}} function performs a final pattern match on its list
argument:
\begin{hscode}\SaveRestoreHook
\column{B}{@{}>{\hspre}l<{\hspost}@{}}%
\column{7}{@{}>{\hspre}l<{\hspost}@{}}%
\column{11}{@{}>{\hspre}l<{\hspost}@{}}%
\column{13}{@{}>{\hspre}l<{\hspost}@{}}%
\column{15}{@{}>{\hspre}l<{\hspost}@{}}%
\column{23}{@{}>{\hspre}l<{\hspost}@{}}%
\column{E}{@{}>{\hspre}l<{\hspost}@{}}%
\>[B]{}{\textbf{\codefont Definition}}\;\Varid{take}\;\mathord{\text{\codefont \{}}\Varid{a}\mathord{\text{\codefont \}}}\mathbin{:}\Conid{Int}\mathbin{\text{\codefont ->}}\Varid{list}\;\Varid{a}\mathbin{\text{\codefont ->}}\Varid{list}\;\Varid{a}\mathbin{:=}{}\<[E]%
\\[-0.3ex]%
\>[B]{}{\textbf{\codefont fix}}\;\Varid{take}\;\Varid{arg\char95 10\char95 \char95 }\;\Varid{arg\char95 11\char95 \char95 }{}\<[E]%
\\[-0.3ex]%
\>[B]{}\hsindent{7}{}\<[7]%
\>[7]{}\mathbin{:=}\textbf{\codefont let}\;\Varid{j\char95 13\char95 \char95 }\mathbin{:=}{}\<[E]%
\\[-0.3ex]%
\>[7]{}\hsindent{4}{}\<[11]%
\>[11]{}{\textbf{\codefont match}}\;\Varid{arg\char95 10\char95 \char95 },\Varid{arg\char95 11\char95 \char95 }\;{\textbf{\codefont with}}{}\<[E]%
\\[-0.3ex]%
\>[11]{}\hsindent{2}{}\<[13]%
\>[13]{}\mid \anonymous ,\Varid{nil}\mathbin{\text{\codefont =>}}\Varid{nil}{}\<[E]%
\\[-0.3ex]%
\>[11]{}\hsindent{2}{}\<[13]%
\>[13]{}\mid \Varid{n},\Varid{cons}\;\Varid{x}\;\Varid{xs}\mathbin{\text{\codefont =>}}{}\<[E]%
\\[-0.3ex]%
\>[13]{}\hsindent{2}{}\<[15]%
\>[15]{}\Varid{cons}\;\Varid{x}\;(\Varid{take}\;(\Varid{op\char95 zm\char95 \char95 }\;\Varid{n}\;(\Varid{fromInteger}\;\mathrm{1}))\;\Varid{xs}){}\<[E]%
\\[-0.3ex]%
\>[7]{}\hsindent{4}{}\<[11]%
\>[11]{}{\textbf{\codefont end}}\;\textbf{\codefont in}{}\<[E]%
\\[-0.3ex]%
\>[7]{}\hsindent{4}{}\<[11]%
\>[11]{}{\textbf{\codefont match}}\;\Varid{arg\char95 10\char95 \char95 },\Varid{arg\char95 11\char95 \char95 }\;{\textbf{\codefont with}}{}\<[E]%
\\[-0.3ex]%
\>[11]{}\hsindent{2}{}\<[13]%
\>[13]{}\mid \Varid{n},\anonymous \mathbin{\text{\codefont =>}}\textbf{\codefont if}\;\Varid{op\char95 zlze\char95 \char95 }\;\Varid{n}\;(\Varid{fromInteger}\;\mathrm{0}){}\<[E]%
\\[-0.3ex]%
\>[13]{}\hsindent{10}{}\<[23]%
\>[23]{}\textbf{\codefont then}\;\Varid{nil}{}\<[E]%
\\[-0.3ex]%
\>[13]{}\hsindent{10}{}\<[23]%
\>[23]{}\textbf{\codefont else}\;\Varid{j\char95 13\char95 \char95 }{}\<[E]%
\\[-0.3ex]%
\>[7]{}\hsindent{4}{}\<[11]%
\>[11]{}{\textbf{\codefont end}}\mathop{\text{.}}{}\<[E]%
\ColumnHook
\end{hscode}\resethooks
The basic idea is to combine multiple equations into a single \ensuremath{{\textbf{\codefont match}}} statement,
whenever possible. We bind these \ensuremath{{\textbf{\codefont match}}} expressions to a name, here \ensuremath{\Varid{j\char95 13\char95 \char95 }},
that earlier patterns return upon pattern failure. We cannot inline this definition, as it would move expressions past the pattern of the second \ensuremath{{\textbf{\codefont match}}} expression, which can lead to unwanted variable capture.

In general, patterns are translated as follows:
\begin{enumerate}
\item We split the alternatives into \emph{mutually exclusive} groups.
 We consider an alternative $\metavar a_1$ to be exclusive with $\metavar a_2$
 if $a_1$ cannot fall through to $a_2$. This is the case if
\begin{enumerate}
\item $\metavar a_1$ has no guards, or
\item an expression matched by the pattern in $\metavar a_1$ will never be
  matched by the pattern in $\metavar a_2$.
\end{enumerate}

\item Each group turns into a single Coq \ensuremath{{\textbf{\codefont match}}} statement which are bound, in
  reverse order, to a fresh identifier. In this translation, the identifier of
  the next group is used as the \emph{fall-through target}.\label{target}

  The last of these groups has nothing to fall-through to. In obviously total
  Haskell, the fall-through will not be needed. Partial code uses
  \ensuremath{\Varid{patternFailure}} as discussed in \cref{sec:partial}.

  If the patterns of the resulting \ensuremath{{\textbf{\codefont match}}} statement are not complete, we add a
  wild pattern case (using \ensuremath{\anonymous }) that returns the fall-through target of the current
  group.

\item Each alternative within such a group turns into one branch of the
  \ensuremath{{\textbf{\codefont match}}}. We translate nested patterns directly, as the semantics of
  patterns in Coq and Haskell coincide on the subset supported by
  \texttt{hs-to-coq}, which excludes incomplete irrefutable patterns, view
  patterns, and pattern synonyms~\cite{pattern-synonyms}.%

  At this point, a \ensuremath{\textbf{\codefont where}} clause in the Haskell code (which spans multiple
  guards) gets translated to a \ensuremath{\textbf{\codefont let}} that spans all guarded right-hand-sides.

\item Each guarded right-hand-side of one alternative gets again bound, in
  reverse order, to a fresh identifier. The last guard uses the fall-through
  target of the whole mutually exclusive group; the other guards use
  the next guard.

\item The sequence of guards of a guarded right-hand-side are now desugared as
  follows:
\begin{enumerate}
\item A boolean guard \ensuremath{\Varid{expr}} turns into
\begin{hscode}\SaveRestoreHook
\column{B}{@{}>{\hspre}l<{\hspost}@{}}%
\column{10}{@{}>{\hspre}l<{\hspost}@{}}%
\column{16}{@{}>{\hspre}c<{\hspost}@{}}%
\column{16E}{@{}l@{}}%
\column{E}{@{}>{\hspre}l<{\hspost}@{}}%
\>[B]{}\textbf{\codefont if}\;\Varid{expr}\;{}\<[10]%
\>[10]{}\textbf{\codefont then}{}\<[16]%
\>[16]{}\;{\ldots}\;{}\<[16E]%
\\[-0.3ex]%
\>[10]{}\textbf{\codefont else}\;{}\<[16]%
\>[16]{}\Varid{j}\text{\phantom{x}}{}\<[16E]%
\ColumnHook
\end{hscode}\resethooks
\item A pattern guard \ensuremath{\Varid{pat}\leftarrow \Varid{expr}} turns into
\begin{hscode}\SaveRestoreHook
\column{B}{@{}>{\hspre}l<{\hspost}@{}}%
\column{18}{@{}>{\hspre}l<{\hspost}@{}}%
\column{25}{@{}>{\hspre}c<{\hspost}@{}}%
\column{25E}{@{}l@{}}%
\column{29}{@{}>{\hspre}c<{\hspost}@{}}%
\column{29E}{@{}l@{}}%
\column{E}{@{}>{\hspre}l<{\hspost}@{}}%
\>[B]{}{\textbf{\codefont match}}\;\Varid{expr}\;{\textbf{\codefont with}}{}\<[18]%
\>[18]{}\mid \Varid{pat}{}\<[25]%
\>[25]{}\mathbin{\text{\codefont =>}}{}\<[25E]%
\>[29]{}\;{\ldots}\;{}\<[29E]%
\\[-0.3ex]%
\>[18]{}\mid \anonymous {}\<[25]%
\>[25]{}\mathbin{\text{\codefont =>}}{}\<[25E]%
\>[29]{}\Varid{j}\text{\phantom{x}}{}\<[29E]%
\ColumnHook
\end{hscode}\resethooks
\item A let guard turns into a \ensuremath{\textbf{\codefont let}} expression scoping over the remaining guards.
\end{enumerate}
Here, \ensuremath{\;{\ldots}\;} is the translation of the remaining guards or, if none are left,
the actual right-hand side expression, and \ensuremath{\Varid{j}} is the current fall-through
target.
\end{enumerate}

This algorithm is not optimal in the sense of producing the fewest \ensuremath{{\textbf{\codefont match}}}
expressions; for example, a more sophisticated notion of mutual exclusivity
could allow an alternative $\metavar a_1$ even when it has guards, as long as
these guards cannot fail (e.g., pattern guards with complete patterns,
\ensuremath{\textbf{\codefont let}}-guards).  This issue has not yet come up in our test cases.

\subsection{Type Classes and Instances}
\label{sec:typeclasses}

Type classes~\cite{typeclasses} are one of Haskell’s most prominent features,
and their success has inspired other languages to implement this feature,
including Coq~\cite{coq-typeclasses}.
As shown in the \text{\tt successors} case study (\cref{sec:successors}), we use this familial relation to translate Haskell type classes into Coq type classes.

As can be seen in \cref{fig:succs.v}, \texttt{hs-to-coq} lifts the method
definitions out of the \ensuremath{{\textbf{\codefont Instance}}}. While not strictly required there,
this lifting is necessary to allow an instance method to refers to another
method of the same instance.

\paragraph{Superclasses}
Superclass constraints are turned into arguments to the generated
class, and these arguments are marked \emph{implicit}, so that Coq’s type class
resolution mechanism finds the right instance. This can be seen in the
definition of \ensuremath{{\textbf{\codefont Class}}\;\Conid{Monad}} in \cref{class-monad}, where the \ensuremath{\Conid{Applicative}} superclass is an implicit argument to \ensuremath{\Conid{Monad}}.

\paragraph{Default Methods}
Haskell allows a type class to declare methods with a default definition. These
definitions are inserted by the compiler into an instance if it omits them. For
example, the code of the \text{\tt successors} library did not give a definition for
\ensuremath{\Conid{Monad}}'s method \ensuremath{\Varid{return}}, and so GHC will use the default definition \ensuremath{\Varid{return}\mathrel{=}\Varid{pure}}.

Since Coq does not have this feature, \texttt{hs-to-coq} has to remember
the default method’s definition and include it in the \ensuremath{{\textbf{\codefont Instance}}} declarations as
needed. This is how the method \ensuremath{\Varid{instance\char95 Monad\char95 Succs\char95 return\char95 }} in
\cref{fig:succs.v} arose.

\paragraph{Derived Instances}
The Haskell standard provides the ability to \emph{derive} a number of basic
type classes (\ensuremath{\Conid{Eq}}, \ensuremath{\Conid{Ord}}, \ldots): the Haskell compiler can optionally
synthesize whole instances of these type classes.  GHC extends this mechanism to
additional type classes (\ensuremath{\Conid{Functor}}, \ensuremath{\Conid{Foldable}}, \ldots). To translate derived
instances, we simply take the instance declarations synthesized by the compiler
and translate them just as we do for user-provided instances.

\paragraph{Self-referential Instances}
Haskell type class instances are in scope even in their own declaration, and
idiomatic Haskell code makes good use of that. Consider the standard instance
for list equality:

\vbox{\begin{hscode}\SaveRestoreHook
\column{B}{@{}>{\hspre}l<{\hspost}@{}}%
\column{3}{@{}>{\hspre}l<{\hspost}@{}}%
\column{11}{@{}>{\hspre}c<{\hspost}@{}}%
\column{11E}{@{}l@{}}%
\column{15}{@{}>{\hspre}l<{\hspost}@{}}%
\column{23}{@{}>{\hspre}l<{\hspost}@{}}%
\column{E}{@{}>{\hspre}l<{\hspost}@{}}%
\>[B]{}\textbf{\codefont instance}\;\Conid{Eq}\;\Varid{a}\mathbin{\text{\codefont =>}}\Conid{Eq}\;[\mskip1.5mu \Varid{a}\mskip1.5mu]\;\textbf{\codefont where}{}\<[E]%
\\[-0.3ex]%
\>[B]{}\hsindent{3}{}\<[3]%
\>[3]{}[\mskip1.5mu \mskip1.5mu]{}\<[11]%
\>[11]{}\mathbin{\text{\codefont ==}}{}\<[11E]%
\>[15]{}[\mskip1.5mu \mskip1.5mu]{}\<[23]%
\>[23]{}\mathrel{=}\Conid{True}{}\<[E]%
\\[-0.3ex]%
\>[B]{}\hsindent{3}{}\<[3]%
\>[3]{}(\Varid{x}\mathbin{:}\Varid{xs}){}\<[11]%
\>[11]{}\mathbin{\text{\codefont ==}}{}\<[11E]%
\>[15]{}(\Varid{y}\mathbin{:}\Varid{ys}){}\<[23]%
\>[23]{}\mathrel{=}\Varid{x}\mathbin{\text{\codefont ==}}\Varid{y}\mathbin{\text{\codefont \&\&}}\Varid{xs}\mathbin{\text{\codefont ==}}\Varid{ys}{}\<[E]%
\\[-0.3ex]%
\>[B]{}\hsindent{3}{}\<[3]%
\>[3]{}\anonymous {}\<[11]%
\>[11]{}\mathbin{\text{\codefont ==}}{}\<[11E]%
\>[15]{}\anonymous {}\<[23]%
\>[23]{}\mathrel{=}\Conid{False}{}\<[E]%
\\[-0.3ex]%
\>[B]{}\hsindent{3}{}\<[3]%
\>[3]{}\Varid{xs}{}\<[11]%
\>[11]{}\mathbin{\text{\codefont /=}}{}\<[11E]%
\>[15]{}\Varid{ys}{}\<[23]%
\>[23]{}\mathrel{=}\text{\codefont not}\;(\Varid{xs}\mathbin{\text{\codefont ==}}\Varid{ys}){}\<[E]%
\ColumnHook
\end{hscode}\resethooks
}%
The operator \ensuremath{\mathbin{\text{\codefont ==}}} occurs three times on the right hand side of method
definitions, and all three occurrences have to be treated differently:
\begin{enumerate}
\item In \ensuremath{\Varid{x}\mathbin{\text{\codefont ==}}\Varid{y}}, which compares list elements, we want to use the
  polymorphic \ensuremath{\Varid{op\char95 zeze\char95 \char95 }} method, so that Coq’s instance resolution 
  picks up the instance for \ensuremath{\Conid{Eq}\;\Varid{a}}.
\item For the first \ensuremath{\Varid{xs}\mathbin{\text{\codefont ==}}\Varid{ys}}, where lists are compared, we cannot use the
  polymorphic method, because the instance \ensuremath{\Conid{Eq}\;[\mskip1.5mu \Varid{a}\mskip1.5mu]} is not yet in
  scope. Instead, we want to refer to the very function that we are defining,
  so we have to turn that function into a fixed point.
\item The second \ensuremath{\Varid{xs}\mathbin{\text{\codefont ==}}\Varid{ys}}, in the definition of \ensuremath{\mathbin{\text{\codefont /=}}}, also cannot be the
  polymorphic method. Instead, we want to refer to the method function for
  list equality that we have just defined.
\end{enumerate}
Unfortunately, \texttt{hs-to-coq} does not have the necessary type instance
resolution information to reliably detect which variant to use. Therefore, we
use following heuristic: By default, the polymorphic method is used. But in a
method definition that is generated based on a \emph{default method}, the
currently defined methods are used.  When this heuristic fails (producing code
that Coq does not accept), the user can inject the correct definition using
\EditName{redefine} edits.

\paragraph{Recursion Through Instances}

We found that nested recursion through higher-order type class methods is a
common Haskell idiom.  A typical example is
\begin{hscode}\SaveRestoreHook
\column{B}{@{}>{\hspre}l<{\hspost}@{}}%
\column{E}{@{}>{\hspre}l<{\hspost}@{}}%
\>[B]{}\textbf{\codefont data}\;\Conid{RoseTree}\;\Varid{a}\mathrel{=}\Conid{Node}\;\Varid{a}\;[\mskip1.5mu \Conid{RoseTree}\;\Varid{a}\mskip1.5mu]{}\<[E]%
\\[\blanklineskip]%
\>[B]{}\Varid{allT}\mathbin{::}(\Varid{a}\mathbin{\text{\codefont ->}}\Conid{Bool})\mathbin{\text{\codefont ->}}\Conid{RoseTree}\;\Varid{a}\mathbin{\text{\codefont ->}}\Conid{Bool}{}\<[E]%
\\[-0.3ex]%
\>[B]{}\Varid{allT}\;\Varid{p}\;(\Conid{Node}\;\Varid{x}\;\Varid{ts})\mathrel{=}\Varid{p}\;\Varid{x}\mathbin{\text{\codefont \&\&}}\Varid{all}\;(\Varid{allT}\;\Varid{p})\;\Varid{ts}{}\<[E]%
\ColumnHook
\end{hscode}\resethooks
Here, the recursive call to \ensuremath{\Varid{allT}} occurs (partially applied) as an argument to
\ensuremath{\Varid{all}}; \ensuremath{\Varid{all}} itself is defined in terms of \ensuremath{\Varid{foldMap}}, a method of the \ensuremath{\Conid{Foldable}}
type class.  The \ensuremath{\Conid{Foldable}} instance for the list type then defines \ensuremath{\Varid{foldMap}} in
terms of \ensuremath{\Varid{foldr}}.

With the naive type class translation outlined above, Coq rejects this.
During termination checking, Coq unfolds definitions of called functions, but
not of pattern matched values; thus, it gets stuck when expanding \ensuremath{\Varid{foldr}},
which becomes
\begin{hscode}\SaveRestoreHook
\column{B}{@{}>{\hspre}l<{\hspost}@{}}%
\column{3}{@{}>{\hspre}l<{\hspost}@{}}%
\column{E}{@{}>{\hspre}l<{\hspost}@{}}%
\>[B]{}{\textbf{\codefont match}}\;\Varid{instance\char95 Foldable\char95 list}\;{\textbf{\codefont with}}{}\<[E]%
\\[-0.3ex]%
\>[B]{}\hsindent{3}{}\<[3]%
\>[3]{}\mid \Conid{Build\char95 Foldable}\;{\ldots}\;\Varid{list\char95 foldr}\;{\ldots}\;\mathbin{\text{\codefont =>}}\;{\ldots}\;{}\<[E]%
\\[-0.3ex]%
\>[B]{}{\textbf{\codefont end}}{}\<[E]%
\ColumnHook
\end{hscode}\resethooks
as \ensuremath{\Varid{foldr}} must be extracted from the type class instance.

We circumvent this issue by using the following generally applicable trick:
Instead of the usual class and instance declarations such as
\begin{hscode}\SaveRestoreHook
\column{B}{@{}>{\hspre}l<{\hspost}@{}}%
\column{E}{@{}>{\hspre}l<{\hspost}@{}}%
\>[B]{}{\textbf{\codefont Class}}\;\Conid{C}\;\Varid{a}\mathbin{:=}\mathord{\text{\codefont \{}}\Varid{method}\mathbin{:}\Varid{t}\;\Varid{a}\mathord{\text{\codefont \}}}\mathop{\text{.}}{}\<[E]%
\\[\blanklineskip]%
\>[B]{}{\textbf{\codefont Instance}}\;\Varid{instance\char95 C\char95 T}\mathbin{:}\Conid{C}\;\Conid{T}\mathbin{:=}\mathord{\text{\codefont \{}}\Varid{method}\mathbin{:=}\Varid{e}\mathord{\text{\codefont \}}}{}\<[E]%
\ColumnHook
\end{hscode}\resethooks
we transform the class into continuation-passing style:
\begin{hscode}\SaveRestoreHook
\column{B}{@{}>{\hspre}l<{\hspost}@{}}%
\column{3}{@{}>{\hspre}l<{\hspost}@{}}%
\column{E}{@{}>{\hspre}l<{\hspost}@{}}%
\>[B]{}{\textbf{\codefont Record}}\;\Conid{C\char95 dict}\;\Varid{a}\mathbin{:=}\mathord{\text{\codefont \{}}\Varid{method\char95 }\mathbin{:}\Varid{t}\;\Varid{a}\mathord{\text{\codefont \}}}\mathop{\text{.}}{}\<[E]%
\\[-0.3ex]%
\>[B]{}{\textbf{\codefont Definition}}\;\Conid{C}\;\Varid{a}\mathbin{:=}{\textbf{\codefont forall}}\;\Varid{r},(\Conid{C\char95 dict}\;\Varid{a}\mathbin{\text{\codefont ->}}\Varid{r})\mathbin{\text{\codefont ->}}\Varid{r}\mathop{\text{.}}{}\<[E]%
\\[-0.3ex]%
\>[B]{}{\textbf{\codefont Existing}}\;{\textbf{\codefont Class}}\;\Conid{C}\mathop{\text{.}}{}\<[E]%
\\[-0.3ex]%
\>[B]{}{\textbf{\codefont Definition}}\;\Varid{method}\;\mathord{\text{\codefont \{}}\Varid{a}\mathord{\text{\codefont \}}}\;\mathord{\text{\codefont \{}}\Conid{H}\mathbin{:}\Conid{C}\;\Varid{a}\mathord{\text{\codefont \}}}\mathbin{:}\Varid{t}\;\Varid{a}{}\<[E]%
\\[-0.3ex]%
\>[B]{}\hsindent{3}{}\<[3]%
\>[3]{}\mathbin{:=}\Conid{H}\;\anonymous \;(\Varid{method\char95 }\;\Varid{a})\mathop{\text{.}}{}\<[E]%
\\[\blanklineskip]%
\>[B]{}{\textbf{\codefont Instance}}\;\Varid{instance\char95 C\char95 T}\mathbin{:}\Conid{C}\;\Conid{T}\mathbin{:=}{}\<[E]%
\\[-0.3ex]%
\>[B]{}\hsindent{3}{}\<[3]%
\>[3]{}{\textbf{\codefont fun}}\;\anonymous \;\Varid{k}\mathbin{\text{\codefont =>}}\Varid{k}\;\mathord{\text{\codefont \{}}\!\!\mid \Varid{method\char95 }\mathbin{:=}\Varid{e}\mid \!\!\mathord{\text{\codefont \}}}\mathop{\text{.}}{}\<[E]%
\ColumnHook
\end{hscode}\resethooks
This neither changes the types of the class methods nor affects instance
resolution, so existing code that uses the type class does not need to be
modified.
Now all method \emph{and instance} definitions are functions, which allows the
termination checker to look through them and accept recursive definitions, such
as \ensuremath{\Varid{allT}}, as structurally recursive.

\subsection{Order of Declarations}
\label{sec:order}

In Haskell, the order of declarations in a source file is irrelevant; functions,
types, type classes, and instances can be used before they are defined. Haskell
programmers often make use of this feature.  Coq, however, requires declarations
to precede their uses.
In order to appease Coq, \texttt{hs-to-coq} detects the dependencies between the
sentences of the Coq file -- a sentence that uses a name depends on it -- and
uses this to sort the sentences topologically so that definitions precede uses.
Mutual recursion is currently unsupported, although this technique naturally
generalizes to include it by treating mutually-recursive groups as a single
node.

While this works in most cases, due to the desugaring of type class constraints
as invisible implicit arguments (\cref{sec:typeclasses}), this process does not
always yield the correct order. In such cases, the user can declare
additional dependencies between definitions by adding an \EditName{order} like
\begin{myhscode}%
\defaultcolumn{@{}>{\codefont}l@{\ }}%
\> {\codefont\textbf{order}} instance\_Functor\_Dual instance\_Monad\_Dual
\end{myhscode}
to the edit file.

\subsection{Partial Haskell}
\label{sec:partial}

Another feature\footnote{Some might prefer quotes around this word.} of
Haskell is that it permits partial functions and general recursion. We
have only discussed verifying \emph{total} Haskell. Nevertheless, as
one starts to work on an existing or evolving Haskell code base, making every
function total and obviously terminating should not have to be the first step.

Therefore, \texttt{hs-to-coq} takes liberties to produce something
useful, rather than refusing to translate partial functions. This way,
verification can already start and inform further development of the Haskell
code. When the design stabilizes, the code can be edited for making
totality obvious.

We can classify translation problems into four categories:
\begin{enumerate}
\item Haskell code with genuinely partial pattern matches; for example,
\begin{hscode}\SaveRestoreHook
\column{B}{@{}>{\hspre}l<{\hspost}@{}}%
\column{E}{@{}>{\hspre}l<{\hspost}@{}}%
\>[B]{}\Varid{head}\mathbin{::}[\mskip1.5mu \Varid{a}\mskip1.5mu]\mathbin{\text{\codefont ->}}\Varid{a}{}\<[E]%
\\[-0.3ex]%
\>[B]{}\Varid{head}\;(\Varid{x}\mathbin{:}\anonymous )\mathrel{=}\Varid{x}{}\<[E]%
\ColumnHook
\end{hscode}\resethooks
  which will crash when passed an empty list.

\item Haskell code with pattern matches that look partial, but are total in a
  way that Coq’s totality checker cannot see. For example, we can define a
  run-length encoding function in terms of \ensuremath{\Varid{group}\mathbin{::}\Conid{Eq}\;\Varid{a}\mathbin{\text{\codefont =>}}[\mskip1.5mu \Varid{a}\mskip1.5mu]\mathbin{\text{\codefont ->}}[\mskip1.5mu [\mskip1.5mu \Varid{a}\mskip1.5mu]\mskip1.5mu]}:
\begin{hscode}\SaveRestoreHook
\column{B}{@{}>{\hspre}l<{\hspost}@{}}%
\column{3}{@{}>{\hspre}l<{\hspost}@{}}%
\column{E}{@{}>{\hspre}l<{\hspost}@{}}%
\>[B]{}\Varid{runLengthEncoding}\mathbin{::}\Conid{Eq}\;\Varid{a}\mathbin{\text{\codefont =>}}[\mskip1.5mu \Varid{a}\mskip1.5mu]\mathbin{\text{\codefont ->}}[\mskip1.5mu (\Varid{a},\Conid{Int})\mskip1.5mu]{}\<[E]%
\\[-0.3ex]%
\>[B]{}\Varid{runLengthEncoding}\mathrel{=}{}\<[E]%
\\[-0.3ex]%
\>[B]{}\hsindent{3}{}\<[3]%
\>[3]{}\Varid{map}\;(\mathord{\text{\codefont \textbackslash}} (\Varid{x}\mathbin{:}\Varid{xs})\mathbin{\text{\codefont ->}}(\Varid{x},\mathrm{1}\mathbin{+}\Varid{length}\;\Varid{xs}))\mathop{\text{.}}\Varid{group}{}\<[E]%
\ColumnHook
\end{hscode}\resethooks
  Since the \ensuremath{\Varid{group}} function returns a list of \emph{nonempty} lists, the partial
  pattern in the lambda will actually always match, but this proof is beyond
  Coq's automatic reasoning.

\item Haskell code with genuinely infinite recursion, at least when evaluated
  strictly\asz{forward reference to \cref{sec:coinduction}?}; for example,
\begin{hscode}\SaveRestoreHook
\column{B}{@{}>{\hspre}l<{\hspost}@{}}%
\column{E}{@{}>{\hspre}l<{\hspost}@{}}%
\>[B]{}\Varid{repeat}\mathbin{::}\Varid{a}\mathbin{\text{\codefont ->}}[\mskip1.5mu \Varid{a}\mskip1.5mu]{}\<[E]%
\\[-0.3ex]%
\>[B]{}\Varid{repeat}\;\Varid{x}\mathrel{=}\Varid{x}\mathbin{:}\Varid{repeat}\;\Varid{x}{}\<[E]%
\ColumnHook
\end{hscode}\resethooks
  produces an infinite list in Haskell, but would diverge in Coq using the
  inductive definition of lists.

\item Haskell code with recursion that looks infinite, but terminates in a way
  that Coq’s termination checker cannot see.  For example, we can implement a
  \ensuremath{\Varid{sort}} function in terms of the standard functional quicksort-like algorithm:

\begin{minipage}{\linewidth}
\begin{hscode}\SaveRestoreHook
\column{B}{@{}>{\hspre}l<{\hspost}@{}}%
\column{5}{@{}>{\hspre}l<{\hspost}@{}}%
\column{E}{@{}>{\hspre}l<{\hspost}@{}}%
\>[B]{}\Varid{sort}\mathbin{::}\Conid{Ord}\;\Varid{a}\mathbin{\text{\codefont =>}}[\mskip1.5mu \Varid{a}\mskip1.5mu]\mathbin{\text{\codefont ->}}[\mskip1.5mu \Varid{a}\mskip1.5mu]{}\<[E]%
\\[-0.3ex]%
\>[B]{}\Varid{sort}\;[\mskip1.5mu \mskip1.5mu]\mathrel{=}[\mskip1.5mu \mskip1.5mu]{}\<[E]%
\\[-0.3ex]%
\>[B]{}\Varid{sort}\;(\Varid{p}\mathbin{:}\Varid{xs})\mathrel{=}\Varid{sort}\;\Varid{lesser}\mathbin{\text{\codefont ++}}[\mskip1.5mu \Varid{p}\mskip1.5mu]\mathbin{\text{\codefont ++}}\Varid{sort}\;\Varid{greater}{}\<[E]%
\\[-0.3ex]%
\>[B]{}\hsindent{5}{}\<[5]%
\>[5]{}\textbf{\codefont where}\;(\Varid{lesser},\Varid{greater})\mathrel{=}\Varid{partition}\;(\mathbin{<}\Varid{p})\;\Varid{xs}{}\<[E]%
\ColumnHook
\end{hscode}\resethooks
\end{minipage}
This function recurses on two lists that are always smaller than the argument,
but not syntactically, so it would be rejected by Coq's termination checker.

\end{enumerate}

Our tool  recognizes partial pattern matches, as described in
\cref{sec:guards}. If these occur, it adds the axiom
\begin{hscode}\SaveRestoreHook
\column{B}{@{}>{\hspre}l<{\hspost}@{}}%
\column{E}{@{}>{\hspre}l<{\hspost}@{}}%
\>[B]{}{\textbf{\codefont Local}}\;{\textbf{\codefont Axiom}}\;\Varid{patternFailure}\mathbin{:}{\textbf{\codefont forall}}\;\mathord{\text{\codefont \{}}\Varid{a}\mathord{\text{\codefont \}}},\Varid{a}\mathop{\text{.}}{}\<[E]%
\ColumnHook
\end{hscode}\resethooks
to the output and completes the pattern match with it, e.g.:
\vbox{\begin{hscode}\SaveRestoreHook
\column{B}{@{}>{\hspre}l<{\hspost}@{}}%
\column{3}{@{}>{\hspre}l<{\hspost}@{}}%
\column{20}{@{}>{\hspre}l<{\hspost}@{}}%
\column{32}{@{}>{\hspre}l<{\hspost}@{}}%
\column{E}{@{}>{\hspre}l<{\hspost}@{}}%
\>[B]{}{\textbf{\codefont Definition}}\;\Varid{head}\;\mathord{\text{\codefont \{}}\Varid{a}\mathord{\text{\codefont \}}}\mathbin{:}\Varid{list}\;\Varid{a}\mathbin{\text{\codefont ->}}\Varid{a}\mathbin{:=}{}\<[E]%
\\[-0.3ex]%
\>[B]{}\hsindent{3}{}\<[3]%
\>[3]{}{\textbf{\codefont fun}}\;\Varid{arg\char95 10\char95 \char95 }\mathbin{\text{\codefont =>}}{}\<[20]%
\>[20]{}{\textbf{\codefont match}}\;\Varid{arg\char95 10\char95 \char95 }\;{\textbf{\codefont with}}{}\<[E]%
\\[-0.3ex]%
\>[20]{}\mid \Varid{cons}\;\Varid{x}\;\anonymous {}\<[32]%
\>[32]{}\mathbin{\text{\codefont =>}}\Varid{x}{}\<[E]%
\\[-0.3ex]%
\>[20]{}\mid \anonymous {}\<[32]%
\>[32]{}\mathbin{\text{\codefont =>}}\Varid{patternFailure}{}\<[E]%
\\[-0.3ex]%
\>[20]{}{\textbf{\codefont end}}\mathop{\text{.}}{}\<[E]%
\ColumnHook
\end{hscode}\resethooks
}
Naturally, this axiom is glaringly unsound. But it does allow the user to
continue translating and proving, and to revisit this issue at a more convenient time --
for example, when they are confident that the overall structure of their project
has stabilized. In the case of genuinely partial functions, the user might want
to change their type to be more precise, as we did in \cref{sec:compiler}. In
the case of only superficially partial code like \ensuremath{\Varid{runLengthEncoding}}, small,
local changes to the code may avoid the problem. At any time, the user can use
Coq’s \ensuremath{\Conid{Print}\;\Conid{Assumptions}} command to check if any provisional axioms are left.


For non-structural recursion, we follow a similar path. Since
\texttt{hs-to-coq} itself does not perform termination checking, it translates
all recursive definitions to Coq fixpoints, which must be structurally
recursive. If this causes Coq to reject valid code, the user can use an edit of
the form {\codefont \textbf{nonterminating} sort} to instruct \texttt{hs-to-coq}
to use the following axiom to implement the recursion:
\begin{hscode}\SaveRestoreHook
\column{B}{@{}>{\hspre}l<{\hspost}@{}}%
\column{E}{@{}>{\hspre}l<{\hspost}@{}}%
\>[B]{}{\textbf{\codefont Local}}\;{\textbf{\codefont Axiom}}\;\Varid{unsafeFix}\mathbin{:}{\textbf{\codefont forall}}\;\mathord{\text{\codefont \{}}\Varid{a}\mathord{\text{\codefont \}}},(\Varid{a}\mathbin{\text{\codefont ->}}\Varid{a})\mathbin{\text{\codefont ->}}\Varid{a}\mathop{\text{.}}{}\<[E]%
\ColumnHook
\end{hscode}\resethooks

Again, this axiom is unsound, but allows the programmer to proceed. In fact,
after including the computation axiom
\begin{hscode}\SaveRestoreHook
\column{B}{@{}>{\hspre}l<{\hspost}@{}}%
\column{3}{@{}>{\hspre}l<{\hspost}@{}}%
\column{E}{@{}>{\hspre}l<{\hspost}@{}}%
\>[B]{}{\textbf{\codefont Axiom}}\;\Varid{unroll\char95 unsafeFix}\mathbin{:}{\textbf{\codefont forall}}\;\Varid{a}\;(\Varid{f}\mathbin{:}\Varid{a}\mathbin{\text{\codefont ->}}\Varid{a}),{}\<[E]%
\\[-0.3ex]%
\>[B]{}\hsindent{3}{}\<[3]%
\>[3]{}\Varid{unsafeFix}\;\Varid{f}\mathrel{=}\Varid{f}\;(\Varid{unsafeFix}\;\Varid{f})\mathop{\text{.}}{}\<[E]%
\ColumnHook
\end{hscode}\resethooks
in the file with the proofs, we were able to verify the partial correctness of
the \ensuremath{\Varid{sort}} function above (i.e., if the equation for \ensuremath{\Varid{sort}} indeed has a fixed
point, as per the above axioms, then it always terminates and produces a sorted
version of the input list).

\sloppy

Eventually, though, the user will have to address this issue 
to consider their proofs complete.  They have many options:
\begin{compactitem}
\item They can apply the
common Coq idiom of adding “fuel”: an additional argument that is structurally
decreasing in each iteration.
\item They can tell \text{\tt hs\char45{}to\char45{}coq} to define this function with \ensuremath{{\textbf{\codefont Program}}\;{\textbf{\codefont Fixpoint}}}, using the \EditName{termination} edit to indicate the termination argument and proof.
\item They can replace the translated definition with a handwritten Coq
  definition. The aforementioned \ensuremath{{\textbf{\codefont Program}}\;{\textbf{\codefont Fixpoint}}} command, the \ensuremath{{\textbf{\codefont Function}}}
  command~\cite{coq-function}, and the Equations package~\cite{equations} can
  all be useful for this, as they allow explicit termination proofs using
  measures or well-founded relations.
\item Or, of course, they can refactor the code to
avoid the problematic functions at all.
\end{compactitem}

\fussy

Thus, the intended workflow around partiality and general recursion is to
begin with axioms in place, which is not an unusual approach to proof
development, and eliminate them at the end as necessary.  For example, the
correctness theorem about Hutton’s razor in \cref{sec:compiler}
goes through even \emph{before} changing the \ensuremath{\Varid{exec}} function to avoid the
partial pattern match!  The reason is that the correctness theorem happens to
only make a statement about programs and stacks that do \emph{not} trigger the
pattern match failure.

\subsection{Infinite Data Structures}
\label{sec:coinduction}

As a consequence of Haskell’s lazy evaluation, Haskell data types are inherently
\emph{coinductive}. For example, a value of type \ensuremath{[\mskip1.5mu \Conid{Int}\mskip1.5mu]} can be an infinite
list. This raises the question of whether we should be making use of Coq’s
support for coinductive constructions, and using \ensuremath{{\textbf{\codefont CoInductive}}} instead of
\ensuremath{{\textbf{\codefont Inductive}}} in the translation of Haskell data types.  The two solutions have
real tradeoffs: with corecursion, we would gain the ability to translate
corecursive functions such as \ensuremath{\Varid{repeat}} (mentioned in \cref{sec:partial}) using
\ensuremath{{\textbf{\codefont cofix}}}, but at the price of our present ability to translate recursive
functions such as \ensuremath{\Varid{filter}} and \ensuremath{\Varid{length}}.

We conjecture, based on our experience as Haskell programmers, that there is a
lot of Haskell code that works largely with finite values.
Moreover, many idioms that do use infinite data structures (e.g., \ensuremath{\Varid{zipWith}\;[\mskip1.5mu \mathrm{0}\mathinner{\ldotp\ldotp}\mskip1.5mu]}) can be rewritten to work only with finite values.
And reasoning about coinduction and corecursion is much trickier than reasoning
about induction and recursion, especially in Coq.

\subsection{Unsupported Language Features}
\label{sec:unsupported}

There are language constructs that \texttt{hs-to-coq} simply does not yet
support, such as mutually recursive definitions, incomplete irrefutable
patterns, and a number of language extensions.  If \texttt{hs-to-coq} detects these, then it outputs an axiom with the name and type
of the problematic definition and an explanatory comment, so that it does not hold up the translation of code using this function.

\section{GHC's \text{\tt base} Library}\label{sec:base}

The case studies in \cref{sec:case-studies} build upon a Coq version of GHC's
\text{\tt base} library~\cite{base-4.9.1.0} that we are developing as part of this
project. This section discusses the design questions raised by constructing
such a library. This process also stress-tests
\texttt{hs-to-coq} itself.

\subsection{What is in the Library?}\label{sec:what-is-in-base}

\begin{figure}
\abovedisplayskip=0pt
\belowdisplayskip=0pt
\raggedright
\begin{description}
\item [Primitive types and operations] \hfill \\
  \ensuremath{\Conid{\Conid{GHC}.Prim}}$^\ast$,
  \ensuremath{\Conid{\Conid{GHC}.Tuple}}$^\ast$,
  \ensuremath{\Conid{\Conid{GHC}.Num}}$^\ast$,
  \ensuremath{\Conid{\Conid{GHC}.Char}}$^\ast$,
  \ensuremath{\Conid{\Conid{GHC}.Base}}
\item [Prelude types and classes] \hfill \\
  \ensuremath{\Conid{\Conid{GHC}.Real}}$^\ast$,
  \ensuremath{\Conid{\Conid{GHC}.Enum}}$^\ast$,
  \ensuremath{\Conid{\Conid{Data}.Bits}}$^\ast$,
  \ensuremath{\Conid{\Conid{Data}.Bool}},
  \ensuremath{\Conid{\Conid{Data}.Tuple}},
  \ensuremath{\Conid{\Conid{Data}.Maybe}},
  \ensuremath{\Conid{\Conid{Data}.Either}},
  \ensuremath{\Conid{\Conid{Data}.Void}},
  \ensuremath{\Varid{Data.Function}},
  \ensuremath{\Conid{\Conid{Data}.Ord}}
\item [List operations] \hfill \\
  \ensuremath{\Conid{\Conid{GHC}.List}},
  \ensuremath{\Conid{\Conid{Data}.List}},
  \ensuremath{\Conid{\Conid{Data}.OldList}}
\item [Algebraic structures] \hfill \\
  \ensuremath{\Conid{\Conid{Data}.Functor}},
  \ensuremath{\Conid{\Conid{Data}.\Conid{Functor}.Const}}$^\ast$,
  \ensuremath{\Conid{\Conid{Data}.\Conid{Functor}.Identity}},
  \ensuremath{\Conid{\Conid{Data}.\Conid{Functor}.Classes}},
  \ensuremath{\Conid{\Conid{Control}.Applicative}},
  \ensuremath{\Conid{\Conid{Control}.Monad}},
  \ensuremath{\Conid{\Conid{Control}.\Conid{Monad}.Fail}},
  \ensuremath{\Conid{\Conid{Data}.Monoid}},
  \ensuremath{\Conid{\Conid{Data}.Traversable}},
  \ensuremath{\Conid{\Conid{Data}.Foldable}},
  \ensuremath{\Conid{\Conid{Control}.Category}},
  \ensuremath{\Conid{\Conid{Control}.Arrow}},
  \ensuremath{\Conid{\Conid{Data}.Bifunctor}}
\end{description}
\caption{Coq \text{\tt base} library modules (starred modules are handwritten, all
  others are translated)}
\label{fig:base-modules}
\end{figure}

Our \text{\tt base} library consists of a number of different modules as shown in
Figure~\ref{fig:base-modules}. These modules include definitions of primitive
types (\ensuremath{\Conid{Int}}, \ensuremath{\Conid{Integer}}, \ensuremath{\Conid{Char}}, \ensuremath{\Conid{Word}}) and their primitive operations, and
common data types (\ensuremath{[\mskip1.5mu \mskip1.5mu]}, \ensuremath{\Conid{Bool}}, \ensuremath{\Conid{Maybe}}, \ensuremath{\Conid{Either}}, \ensuremath{\Conid{Void}}, \ensuremath{\Conid{Ordering}},
tuples) and their operations from the standard prelude. They also include
prelude type classes (\ensuremath{\Conid{Eq}}, \ensuremath{\Conid{Ord}}, \ensuremath{\Conid{Enum}}, \ensuremath{\Conid{Bounded}}) as well as classes for
algebraic structures (\ensuremath{\Conid{Monoid}}, \ensuremath{\Conid{Functor}}, \ensuremath{\Conid{Applicative}}, \ensuremath{\Conid{Monad}}, \ensuremath{\Conid{Arrow}},
\ensuremath{\Conid{Category}}, \ensuremath{\Conid{Foldable}}, \ensuremath{\Conid{Traversable}}) and data types that assist with these
instances.

During the development of this library we faced the design decision of whether
we should translate all Haskell code to new Coq definitions, or whether we
should connect Haskell types and functions to parts of the Coq standard library.
We have chosen to do the latter, mapping basic Haskell types (such as \ensuremath{\Conid{Bool}},
\ensuremath{[\mskip1.5mu \mskip1.5mu]}, \ensuremath{\Conid{Maybe}}, and \ensuremath{\Conid{Ordering}}) to their Coq counterparts (respectively
\ensuremath{\Varid{bool}}, \ensuremath{\Varid{list}}, \ensuremath{\Varid{option}}, and \ensuremath{\Varid{comparison}}). This makes the output
slightly less recognizable to Haskell programmers -- users must
know how these types and constructors match up. However, it also makes
existing Coq proofs about these data structures available.

Support for this mapping in \texttt{hs-to-coq} is provided via \EditName{rename} edits,
which allow us to make that decision on a per-type and per-function basis, as
the following excerpt of the edit file shows:
\begin{myhscode}%
\defaultcolumn{@{}>{\codefont}l@{\ }}%
\> \codefont\textbf{rename type}   \>GHC.Types.[]  \>= list \\
\> \codefont\textbf{rename value}  \>GHC.Types.[]  \>= nil \\
\> \codefont\textbf{rename value}  \>GHC.Types.:   \>= cons
\end{myhscode}

The library also includes (handwritten) modules that specify and prove
properties of this code, including type classes that describe \emph{lawful}
functors, applicative functors, and monads, as discussed in
\cref{sec:successors}.  We include proofs that the type constructors \ensuremath{\Varid{list}}
and \ensuremath{\Varid{option}} are lawful functors, applicative functors, and monads by
instantiating these classes.

\subsection{How Did We Develop the Library?}
Because of the nature of this library, some modules are more amenable to automatic
translation than others.  We defined most of the modules via
automatic translation from the GHC source (with the assistance of edit
instructions). The remainder were handwritten, often derived via modification
of the output of automatic translation.

We were forced to manually define modules that define primitive types, such as
\ensuremath{\Conid{\Conid{GHC}.Word}}, \ensuremath{\Conid{\Conid{GHC}.Char}}, and \ensuremath{\Conid{\Conid{GHC}.Num}}, because they draw heavily on a feature
that Coq does not support: unboxed types. Instead, we translate primitive
numeric types to signed and unsigned binary numbers in Coq (\ensuremath{\Conid{Z}} and \ensuremath{\Conid{N}},
respectively). Similarly, we translate \ensuremath{\Conid{Rational}} to Coq's type \ensuremath{\Conid{Q}} of
rational numbers.\footnote{In the case of fixed precision types, we have
  chosen these mappings for expediency; in future work, we plan to switch
  these definitions so that we can reason about underflow and overflow.}
Modules that make extensive use of these primitive types, such as \ensuremath{\Conid{\Conid{GHC}.Enum}}
and \ensuremath{\Conid{\Conid{Data}.Bits}} were also handwritten.  Finally, one module
(\ensuremath{\Conid{\Conid{Data}.\Conid{Functor}.Const}}) was handwritten because it uses features that are
beyond the current scope of our tool.

\sloppy

On the other hand, we are able to successfully generate several modules in the
base library, including the primary file \ensuremath{\Conid{\Conid{GHC}.Base}} and the list libraries
\ensuremath{\Conid{\Conid{GHC}.List}} and \ensuremath{\Conid{\Conid{GHC}.OldList}}. Other notable successes include translating the
algebraic structure libraries \ensuremath{\Conid{\Conid{Data}.Monoid}}, \ensuremath{\Conid{\Conid{Data}.Foldable}},
\ensuremath{\Conid{\Conid{Data}.Traversable}}, and \ensuremath{\Conid{\Conid{Control}.Monad}}.
\scw{We don't talk about safe coercions anywhere. Yet that ability is really
  important for \ensuremath{\Conid{\Conid{Data}.Monoid}}, for example.}

\fussy

Translating these modules requires several forms of edits. As we describe
below, some of these edits are to \EditName{skip} definitions that we do not
wish to translate. We also use edits to augment the translation with
additional information, in order to make the Coq output type check. For
example, these include annotations on the kinds of higher-order datatype
parameters or explicit type instantiations.  Other \EditName{redefine} edits
are necessary to patch up type class instances when the heuristics described
in \cref{sec:typeclasses} fail. Still others are necessary to reorder
definitions, as described in \cref{sec:order}.

\subsection{What is Skipped?}
During the translation process, the edits allow us to \EditName{skip}
Haskell definitions. Most modules had at least one skipped
definition, and under a quarter had more than twenty.

Many of the skipped definitions are due to partiality. For example, we do not
translate functions that could trigger pattern match failure, such as \ensuremath{\Varid{head}} or
\ensuremath{\Varid{maximum}}, or that could diverge, such as \ensuremath{\Varid{repeat}} or \ensuremath{\Varid{iterate}}.

Some type classes have methods that are often instantiated with partial
functions.  We also removed such members, such as the \ensuremath{\Varid{fail}} method of the
\ensuremath{\Conid{Monad}} class (as mentioned in \cref{sec:successors}), the \ensuremath{\Varid{foldl1}},
\ensuremath{\Varid{foldr1}}, \ensuremath{\Varid{maximum}} and \ensuremath{\Varid{minimum}} methods of the \ensuremath{\Conid{Foldable}} class, and the
\ensuremath{\Varid{enumFromThen}} and \ensuremath{\Varid{enumFromThenTo}} methods of the \ensuremath{\Conid{Enum}} class.  In the last case, this
is not \emph{all} of the partial methods of the class; for example, the \ensuremath{\Varid{pred}}
and \ensuremath{\Varid{succ}} methods throw errors in instances for bounded types, and the
\ensuremath{\Varid{enumFrom}} method diverges for infinite types. To solve this problem, we have
chosen to support the \ensuremath{\Conid{Enum}} class only for bounded types. In this case, we
modified the \ensuremath{\Varid{pred}} and \ensuremath{\Varid{succ}} methods so that they return the \ensuremath{\Varid{minBound}} and
\ensuremath{\Varid{maxBound}} elements, respectively, at the end of their ranges. For \ensuremath{\Varid{enumFrom}},
we use \ensuremath{\Varid{maxBound}} to provide an end point of the enumeration.

Some functions are total, but it is difficult for Coq to determine that they
are. For example, the \ensuremath{\Varid{eftInt}} function in the \ensuremath{\Conid{Enum}} module enumerates a list
of integers from a starting number \ensuremath{\Varid{x}} to an ending number \ensuremath{\Varid{y}}. This function is
not structurally recursive, so we use the \ensuremath{{\textbf{\codefont Program}}\;{\textbf{\codefont Fixpoint}}} extension to
provide its termination proof in our redefinition.

Some parts of these modules are skipped because they relate to operations that
are out of scope for our tool. We do not translate any definitions or instances
related to \ensuremath{\Conid{IO}}, so we skip all functionality related to \ensuremath{\Conid{Read}} and \ensuremath{\Conid{Show}}. We
also do not plan to support reflection, so we skip all instances related to
\ensuremath{\Conid{\Conid{GHC}.Generics}}. Similarly, we do not include arrays, so we skip
instances related to array types and indexing.

\section{Related Work}
\label{sec:related}

\sloppy

\paragraph*{Advanced Function Definitions}
Translating Haskell idioms into Coq pushes the limits of the standard vernacular
to define functions, especially when it comes to the expressiveness of pattern
matching (see \cref{sec:guards}) and non-structural recursion (see
\cref{sec:partial}). A number of Coq extensions aim to alleviate these
limitations:
\begin{itemize}
\item The \ensuremath{{\textbf{\codefont Program}}\;{\textbf{\codefont Fixpoint}}} and \ensuremath{{\textbf{\codefont Function}}} vernacular commands, which are part
  of the Coq distribution, permit non-structural recursion by specifying a
  decreasing termination measure or a well-founded relation. We found that
  \ensuremath{{\textbf{\codefont Program}}\;{\textbf{\codefont Fixpoint}}} works better in the presence of nested recursion through
  higher-order functions, and \text{\tt hs\char45{}to\char45{}coq} supports generating \ensuremath{{\textbf{\codefont Program}}\;{\textbf{\codefont Fixpoint}}}
  definitions.
\item The Equations plugin~\cite{equations} provides Coq support for
  well-founded recursion to and Agda-style dependent pattern matching.  It
  supports nested recursion through higher-order functions as least as well as
  \ensuremath{{\textbf{\codefont Program}}\;{\textbf{\codefont Fixpoint}}}, and furthermore produces more usable lemmas (e.g.,
  unfolding equations). However, its changes to the pattern matching syntax,
  while improving support for dependent pattern matching, do not include support
  for guards with fall-through semantics and do not support non-top-level
  \ensuremath{{\textbf{\codefont match}}} expressions, both of which are important for our translation.
\end{itemize}

\fussy

\subsection{Haskell and Coq}

\paragraph*{Extraction}
The semantic proximity of (total) Haskell and Coq, which we rely on, is also
used in the other direction by Coq’s support for code extraction to
Haskell~\cite{code-extraction}. Several projects use this feature to verify
Haskell code~\cite{Chen:2015,Megacz}.  However, since extraction starts with Coq
code and generates Haskell, it cannot be used to verify pre-existing Haskell
code.  Although in a certain sense, \text{\tt hs\char45{}to\char45{}coq} and extraction are inverses,
round-tripping does not produce syntactically equivalent output in either
direction. On the one hand, \text{\tt hs\char45{}to\char45{}coq} extensively annotates the resulting Coq
code; on the other, extraction ignores many Haskell features and inserts unsafe
type coercions. In future work, we hope to use testing to verify that
round-tripping produces operationally equivalent output; this would provide
greater assurance about the correctness of both \texttt{hs-to-coq} and extraction.


\paragraph*{Manual Translation} The \texttt{coq-haskell} library~\cite{coq-haskell} is a
hand-written Coq library designed to make it easier for Haskell programmers to
work in Coq.  In many ways, it serves a similar purpose to our translation of
\text{\tt base} (\cref{sec:base}).  In addition to enabling easier Coq programming, it
also provides support for extracting Coq programs to Haskell.

\subsection{Haskell and First-order Logic}

\paragraph*{LiquidHaskell} LiquidHaskell~\cite{Vazou:2014:RTH:2628136.2628161}
augments the Haskell programming language with refinement types: all types can
be coupled with a predicate that the inhabitants must satisfy.  These
refinements are then automatically checked by an SMT solver; a successful
solution means that all functions are total and conform to these new, richer,
specifications.
In practice, when proving theorems in Coq, we can take
advantage of a mature environment and proof automation techniques; this can allow
for faster verification than LiquidHaskell, given a corresponding Coq
program~\cite{Vazou:2017:TTP:3122955.3122963}.


\paragraph*{Halo} The prototype contract checker \text{\tt halo}~\cite{Vytiniotis:2013}
takes a Haskell program, uses GHC to desugar it into the intermediate language
Core, and then translates the Core program into a first-order logic formula.  It
then invokes an SMT solver such as Z3~\cite{MouraB08} or
Vampire~\cite{RiazanovV99} to prove this formula; a successful proof tells us
that the original program is crash-free.


\subsection{Translating Haskell to Higher-order Logic}

\paragraph*{Haskabelle}
In the Isabelle/HOL ecosystem, \texttt{hs-to-coq} has a direct correspondence in Haskabelle~\cite{haskabelle}, which translates total Haskell code into equivalent Isabelle function definitions. Like our tool, it parses Haskell, desugars syntactic constructs, configurably
adapts basic types and functions to their counterpart in Isabelle’s standard library. It used to be bundled with the Isabelle release, but it has not been updated recently and was dropped from Isabelle.

While Isabelle/HOL is, like Coq, a logic of total functions, all types in HOL are non-empty and
inhabited by the polymorphic value \ensuremath{\text{\codefont undefined}}. Therefore, Haskabelle
can translate partial patterns like described in \cref{sec:guards}, but without
introducing inconsistency by relying on axioms.

Haskabelle supports boolean guards in simple cases, but does not implement
fall-through on guard failure.  In particular, the \ensuremath{\Varid{take}}
function shown in \cref{sec:guards} would be translated to a function that is
\ensuremath{\text{\codefont undefined}} when \ensuremath{\Varid{n}\mathbin{>}\mathrm{0}}.

\paragraph*{HOLCF-Prelude}
A translation of Haskell into a total logic, as performed by \texttt{hs-to-coq}
and Haskabelle, necessarily hides the finer semantic nuances that arise
due to laziness, and does not allow reasoning about partially defined or
infinite values. If that is a concern, one might prefer a translation into the
Logic of Computable Functions (LCF)~\cite{LCF}, where every type is a domain and
every function is continuous.  LCF is, for example, implemented in Isabelle’s
HOLCF package~\cite{HOLCF,HOLCF11}. Parts of the Haskell standard library have
been manually translated into this setting~\cite{holcf-prelude} and used to
verify the rewrite rules applied by \texttt{HLint}, a Haskell style checker.

\paragraph*{seL4}
Haskell has been used as a prototyping language for formally verified systems
in the past. The seL4 verified microkernel started with a Haskell prototype
that was semi-automatically translated to
Isabelle/HOL~\cite{Derrin:2006:RMA}. As in our work, they were restricted to
the terminating fragment of Haskell.

The authors found that the availability of the Haskell prototype provided a
machine-checkable formal executable specification of the system. They used
this prototype to refine their designs via testing, allowing them to make corrections
before full verification. In the end, they found that starting with Haskell led to a
``productive, highly iterative development'' contributing to a ``mature final
design in a shorter period of time.''
\leo{I buy the potential workflow much more now that I read the sel4
  quote. Maybe rework that in the introduction?}

\subsection{Haskell and Dependently-typed Languages}

\paragraph*{Programmatica/Alfa}
The Programmatica project~\cite{Hallgren04anoverview} included a tool to
translate Haskell into the proof editor Alfa.  As in our work, their tool only
produces valid proofs for total functions over finite data structures. They
state: ``When the translation falls outside that set, any correctness proofs
constructed in Alfa entail only partial correctness, and we leave it to the
user to judge the value of such proofs.''

The logic of the Alfa proof assistant is based on dependent type theory, but
without as many features as Coq. In particular, the Programmatica tool compiles away type classes
and nested pattern matching, features retained by \texttt{hs-to-coq}.

\paragraph*{Agda 1}
Dyber, Haiyan, and Takeyama~\cite{DybjerHT04} developed a tool for automatically
translating Haskell programs to the Agda/Alfa proof assistant. Their solution to
the problem of partial pattern matching is to synthesize predicates that
describe the domain of functions.  They explicitly note the
interplay between testing and theorem proving and show how to verify a tautology checker.

\paragraph*{Agda 2}
Abel~et~al.~\cite{Abel:2005} translate Haskell expressions into the logic of the
Agda 2 proof assistant. Their tool works later in the GHC pipeline than ours;
instead of translating Haskell source code, they translate Core expressions.
Core is an explicitly typed internal language for Haskell used by GHC, where type
classes, pattern matching and many forms of syntactic sugar have been compiled
away.

Their translation explicitly handles partiality by introducing a monad for
partial computation. Total code is actually polymorphic over the monad in which
it should execute, allowing the monad to be instantiated by the identity monad
or the \ensuremath{\Conid{Maybe}} monad as necessary. Agda's predicativity also causes issues with
the translation of GHC's impredicative, System F-based core language.

\subsection{Translating Other Languages to Coq}

Chargueraud's CFML~\cite{chargueraud-10-cfml} translates OCaml source code to
characteristic formulae expressed as Coq axioms. This system has been used to
verify many of the functional programs from Okasaki's Purely Functional Data
Structures~\cite{okasaki}.%
%

\section{Conclusions and Future Work}
\label{sec:conclusion}

\sloppy

We presented a methodology for verifying Haskell programs, built
around translating them into Coq with the \texttt{hs-to-coq} tool.  We
successfully applied this methodology to pre-existing code in multiple case
studies, as well as in the ongoing process of providing the \text{\tt base} Haskell
library for these and other examples to build on.

\fussy

Looking forward, there are always more Haskell features that we can extend the
tool to support; we plan to apply this tool to larger real-world software
projects and will use that experience to prioritize our next steps. We also
would like to develop a Coq tactic library that can help automate reasoning
about the patterns found in translated Haskell code as well as extend the
proof theory of our \text{\tt base} library.

\paragraph*{Acknowledgments}
Thanks to John Wiegley for discussion and support, and to Leonidas Lampropoulos
and Jennifer Paykin for their helpful comments.
This material is based upon work supported by the
  \grantsponsor{GS100000001}{National Science
    Foundation}{http://dx.doi.org/10.13039/100000001} under Grant
  No.~\grantnum{GS100000001}{1319880} and Grant
  No.~\grantnum{GS100000001}{1521539}.

\bibliography{bib}

\end{document}